\documentclass[12pt, a4paper]{article}
\pdfoutput=1

\usepackage{amsmath}
\usepackage{amsfonts}
\usepackage{amssymb}
\usepackage{graphicx, rotating}
\usepackage{epsfig}
\usepackage{latexsym}
\usepackage{graphicx}
\usepackage{color}
\usepackage{amsmath,bm,amssymb}
\usepackage{cite}
\usepackage{slashed}
\usepackage{hyperref}
\hypersetup{colorlinks, citecolor=bluscuro, linkcolor=black, urlcolor=bluscuro}
\definecolor{rossos}{cmyk}{0,1,1,0.55}
\definecolor{bluscuro}{rgb}{0.15, 0.2, .85}
\definecolor{bluchiaro}{cmyk}{1,.3,0.,0.1}

\numberwithin{equation}{section}

\newcommand{\be}{\begin{equation}}
\newcommand{\ee}{\end{equation}}
\newcommand{\bea}{\begin{eqnarray}}
\newcommand{\eea}{\end{eqnarray}}

\def\simlt{\stackrel{<}{{}_\sim}}

\def\pp{{\scriptscriptstyle +}}
\def\mm{{\scriptscriptstyle -}}
\def\pms{{\scriptscriptstyle{\pm}}}

\newcommand{\arXiv}[2]{\href{http://arxiv.org/pdf/#1}{{\tt [#2/#1]}}}
\newcommand{\arXivold}[1]{\href{http://arxiv.org/pdf/hep-#1}{{\tt [#1]}}}

\def\bma#1{\mbox{\boldmath{$#1$}}}

\begin{document}
\allowdisplaybreaks
\begin{titlepage}
\begin{flushright}
IFT-UAM/CSIC-20-71
\end{flushright}
\vspace{.3in}

\vspace{1cm}
\begin{center}
{\Large\bf\color{black} 
The Stabilizing Effect of Gravity Made Simple
} \\
\vspace{1cm}{
{\large J.R.~Espinosa}
} \\[7mm]
{\it Instituto de F\'{\i}sica Te\'orica UAM/CSIC, \\ 
C/ Nicol\'as Cabrera 13-15, Campus de Cantoblanco, 28049, Madrid, Spain
}
\end{center}
\bigskip

\vspace{.4cm}

\begin{abstract}
A new approach to vacuum decay in quantum field theory, based on a simple variational formulation in field space using a tunneling potential, is ideally suited to study the effects of gravity on such decays. The method allows to prove in new and simple ways many results, among others, that gravitational corrections tend to make Minkowski or Anti de Sitter false vacua more stable semiclassically or that higher barriers increase vacuum lifetime. The approach also offers a very clean picture of gravitational quenching of vacuum decay and its parametric dependence on the features of a potential and allows to study the BPS domain-walls between vacua in critical cases. Special attention is devoted to supersymmetric potentials and to the discussion of near-critical vacuum decays, for which it is shown how the new method can be usefully applied beyond the thin-wall approximation.  
\end{abstract}
\bigskip

\end{titlepage}

\section{Introduction\label{sec:introduction}}

The effect of gravity on the decay of a false vacuum in quantum field theory was first addressed in the pioneering work of Coleman and De Luccia \cite{CdL}, already 40 years ago. Although most of the key results were obtained in that work, many refinements and developments have been added since then. 
The aim of this paper is to show how an alternative formulation of vacuum decay offers a new and simple way to study the gravitational impact on those decays. This method provides a very direct route to the derivation of such effects, as we illustrate by obtaining some old as well as some new results. 

The new approach to the calculation of tunneling actions was introduced in \cite{E,Eg} as an alternative to the usual Euclidean bounce approach \cite{CdL}. The method formulates the calculation of the tunneling action describing the decay of a false vacuum at $\phi_\pp$ as an elementary variational problem in the following terms: find the ``tunneling potential'' $V_t(\phi)$ that interpolates between $\phi_\pp$ and the basin of the true vacuum and minimizes an action functional, an integral 
in field space of the appropriate action density, that takes the simple form\footnote{For simplicity, we consider throughout a zero value for the nonminimal coupling $\xi$ of the Higgs to the Ricci scalar. For the general formalism with nonzero $\xi$, see \cite{ESM}.}:
\be
 S[V_t]=\frac{6\pi^2}{\kappa^2}\int_{\phi_\pp}^{\phi_0}d\phi\ \frac{(D+V_t')^2}{V_t^2 D}\ ,
\label{SVt}
\ee
where primes denote field derivatives, and
\be
D^2\equiv V_t'{}^2+6\kappa (V-V_t)V_t\ ,
\ee
with $\kappa=1/m_P^2$  ($m_P$ is the reduced Planck mass) and $\phi_0$ is in the basin of the true vacuum $\phi_\mm$ (by convention we take $\phi_\mm>\phi_\pp$). The action thus obtained reproduces the Euclidean bounce result and the method has various advantages that have been discussed elsewhere \cite{E,Eg,EK}. Among other nice properties, it permits a fast and precise numerical determination of the action; it can be modified in order to study decays by thermal fluctuations; it can be applied to obtain potentials that permit a fully analytical solution to the  tunneling problem; it is very efficient to study vacuum decay in multi-field potentials as one is searching for a minimum of the action (rather than a saddle-point, as in the Euclidean approach), etc.

The Euler-Lagrange equation derived from the stationarity of the action (\ref{SVt}) under variations of $V_t$ gives the following ``equation of motion'' (EoM) for $V_t$:
\be
6(V-V_t)\left[V_t''+\kappa\left(
3V-2V_t\right)\right]
+ V_t'\left(4V_t'-3V'\right)=0
\ ,
\label{EoMVt}
\ee
or, in terms of $D$,
\be
D'= \frac{\left(3V'-4\, V_t'  \right)}{6(V-V_t)}D\ .
\label{EoMVtD}
\ee
In the case of the false vacuum being a Minkowski or anti-de Sitter (AdS) vacuum, the cases of interest for this paper, $V_t$ is monotonic with $V_t'\leq 0$ and boundary conditions 
\be
V_t(\phi_+)=V(\phi_+)\ ,\quad\quad  V_t(\phi_0)=V(\phi_0)\ ,
\label{BCsVt}
\ee
with $\phi_0$ [equal to the central value of the bounce, $\phi_B(0)$, in the Euclidean approach] being an unknown to be found when minimizing (\ref{SVt}) and satisfying $\phi_\pp>\phi_0\geq \phi_\mm$. The EoM for $V_t$ also fixes
\be
V_t'(\phi_+)=V'(\phi_+)=0\ ,\quad
V_t'(\phi_0)=\frac34 V'(\phi_0)\ .
\label{BCsVtp}
\ee

In this paper we are mainly concerned on the stabilizing effects of gravity on the decays of Minkowski or AdS false vacua.\footnote{In contrast, dS vacua are always rendered unstable by gravitational corrections.} The starting point of our analysis is the tunneling action $S[V_t]$ in (\ref{SVt}). Using it, Section~\ref{sec:stabk} proves, in a perturbative expansion in powers of $\kappa$, that gravity tends to make such vacua more stable, increasing the value of the tunneling action. While this proof is carried out at ${\cal O}(\kappa)$, Section~\ref{sec:staballk} extends it to all orders in $\kappa$. This general proof is based in comparing the action density of (\ref{SVt}) with its $\kappa=0$ limit. Using similar comparisons, other general results can be obtained. A simple  illustration of the method shows that larger barriers lead to higher tunneling actions, as one would naively expect. 

Section~\ref{sec:quench} addresses gravitational quenching of vacuum decay \cite{CdL}: how gravitational effects can stabilize completely (at least semi-classically) a false vacuum. The tunneling approach gives a very simple description of this effect, which happens when $D^2>0$ cannot be achieved. This simple condition allows to identify the generic features necessary to quench the decay of a given vacuum. It also shows in a straightforward way how to check whether a false vacuum is subcritical (finite tunneling action) or supercritical (infinite tunneling action). The boundary between these types of vacua/potential correspond to critical cases\footnote{For Minkowski false vacua, this boundary corresponds to the so-called great divide \cite{GDivide}.} for which $D\equiv 0$. Decay between critical vacua is not possible but static domain walls separating the vacua can be formed. The criticality condition also leads immediately to the generic form of potentials with critical false vacua. 
 
One particular class of potentials that lead generically to critical false vacua are supersymmetric ones, which are analyzed in Section~\ref{sec:SUSY}. There the tunneling potential approach is generalized to multiple complex scalar fields and a simple generic form for $V_t$ is derived, as well as the expression for the trajectory in field space of the corresponding domain walls. The stability of supersymmetric vacua hols also when the true vacuum is not supersymmetric. All this is illustrated by several concrete examples. 
Then, Section~\ref{sec:dw} shows how to obtain the domain wall profiles when using the tunneling potential method.

When a false vacuum is subcritical but close to being critical, the decay is very suppressed and this near-critical decay can be treated as a small perturbation of the critical one. This is discussed with some examples in Section~\ref{sec:nearcrit}. Typically one expects such decays to be well described by the thin-wall approximation but this is not always the case. Section~\ref{sec:btw} shows how in such cases one can still devise some approximation based on the smallness of $D$.

Finally, Section~\ref{sec:sumout} provides a summary and an outlook for future applications, while Appendix~\ref{sec:app} contains details of the method for  the multi-field case with non-canonical fields, both for real or complex fields (the latter being need for the analysis of supergravity potentials).

\section{Stabilization Effect of Gravity to $\bma{{\cal O}(\kappa)}$\label{sec:stabk}}
To start analyzing the effect of gravity on the decay of Minkowski or AdS false vacua let us first consider cases for which gravitational corrections give only a small effect. In those cases we can expand the tunneling potential, the Euler-Lagrange equation for $V_t$ and the tunneling action to first order in a perturbative expansion in $\kappa$. This is always an expansion in a dimensionless combination $\kappa M^2$, where $M$ is some typical mass scale characterizing the potential. Small gravitational effects are expected when $\kappa M^2\ll 1$. Let us write 
\be
V_t=V_{t 0}+\kappa V_{t 1}+{\cal O}(\kappa^2)\ ,
\ee 
where $V_{t 0}$ is the tunneling potential with no gravity. 
The general expansion of the tunneling action density ${\it s}(\phi)$ can be recast simply as
\be
{\it s}(\phi)={\it s}_{0}(\phi)\left(1+3\kappa 
\frac{V_{t 0}^2}{V_{t 0}'{}^2}\right)
+{\cal O}(\kappa^2)\ ,
\label{sexp}
\ee
up to a total-derivative term, to be discussed below.
The zero-th order term in this expansion is the tunneling action density without gravity \cite{E}
\be
{\it s}_{0}(\phi)=-54\pi^2\frac{(V-V_{t 0})^2}{V_{t 0}'^3}\ .
\label{s0Vt}
\ee
As $V'_{t 0}\leq 0$ \cite{E}, this action density is positive definite.
The zeroth-order Euler-Lagrange equation for $V_{t 0}$ derived from the corresponding action integral is \cite{E}
\be
\label{EoM0}
(4V_{t 0}'-3V')V_{t 0}'=6(V_{t 0}-V)V_{t 0}''\ .
\ee

The second term in (\ref{sexp}) gives the ${\cal O}(\kappa)$ effects of gravity and is also positive, making the action density larger.  Before concluding that gravity increases the tunneling action, one should also consider how gravity modifies the integration interval by changing the end-point $\phi_0$. However, an ${\cal O}(\kappa)$ change in $\phi_0$ modifies the action integral only at higher order as ${\it s}(\phi_0)=0$. 
So we indeed can conclude that gravity tends to make metastable vacua more stable.
Although this is only a perturbative proof at ${\cal O}(\kappa)$, the stabilizing effect of gravity (for Minkowski or AdS vacua) is more general as is proven in the next section. 

We omitted from (\ref{sexp}) a term that, after using the Euler-Lagrange equation for $V_t$, is a total derivative
\be
\delta {\it s}(\phi)=
162\pi^2\kappa\frac{d}{d\phi}\left\{\frac{(V-V_{t 0})^2}{V_{t 0}'^4}\left[V_{t 1}+\frac{(V-V_{t 0})V_{t 0}^2}{V_{t 0}'^2}\right] \right\}\ ,
\label{sexpb}
\ee
 and contributes a boundary term to the integrated tunneling action. 
However this boundary term gives a zero contribution to the action integral: at $\phi_0$ one has $V_{t 0}(\phi_0)=V(\phi_0)$ with
non-zero $V'_{t 0}(\phi_0)$ so that the boundary term vanishes at $\phi_0$. At $\phi_\pp=0$ (without loss of generality we can take $\phi_+=0$) one needs to know how the functions $V, V_{t 0}$ and $V_{t 1}$ approach zero.  
 In order that the integral of the action density (\ref{s0Vt}) does not diverge at $\phi=0$ we should have
\be
\lim_{\phi\rightarrow 0}  \frac{\phi (V-V_{t 0})^2}{V_{t 0}'^3} =0\ .
\label{nodiv}
\ee
To prove that  the term inside the curly brackets in (\ref{sexpb}) goes to zero for $\phi\rightarrow 0$, it is convenient to rewrite it as
\be
\left[\frac{\phi (V-V_{t 0})^2}{V_{t 0}'^3}\right]
\left(\frac{V_{t 0}}{\phi V_{t 0}'}\right)\frac{V_{t 1}}{V_{t 0}}+\left[
\frac{\phi (V-V_{t 0})^2}{V_{t 0}'^3}\right]^{3/2}\left(
\frac{V_{t 0}}{\phi V_{t 0}'}\right)^{3/2}V_{t 0}^{1/2}\ ,
\ee
and to note that the terms in square brackets tend to zero by (\ref{nodiv}); those in parentheses go to a constant;
$V_{t 0}\rightarrow 0$; and $V_{t 1}/V_{t 0}$ should tend to zero or a constant as $V_{t 1}$ cannot go to zero more slowly than $V_{t 0}$. (If it did, gravitational effects would not be small at small $\phi$ and the perturbative expansion would not  be applicable there.)

An interesting property of the expansion result (\ref{sexp}) is that the  ${\cal O}(\kappa)$ term depends only on zero-th order quantities.\footnote{This fact extends to higher order corrections, which depend only on lower order terms and has been used with advantage in \cite{ESM} to study the tunneling action for decay of  the Standard Model vacuum.} This is ultimately due to the fact that we expand the action around its minimum so that a first order shift in $V_t$ changes the action value only at second order.

\section{Stabilization Effect of Gravity to All Orders in $\bma\kappa$\label{sec:staballk}}

The stabilization effect of gravity is general and holds also when gravitational corrections are not small \cite{Quench2}. The tunneling potential approach can be used to give a straightforward proof of this fact.
Consider a given path $V_t(\phi)$ out of the metastable vacuum 
$\phi_\pp$ of a given potential $V(\phi)$. The tunneling action densities with and without gravity satisfy 
\be
{\it s}(V_t)\equiv \frac{6\pi^2}{\kappa^2}\frac{(D+V_t')^2}{D V_t^2}
\geq 
{\it s}_0(V_t)\equiv 54\pi^2\frac{(V-V_t)^2}{-V_t'^3}
\ .\label{skgeqs0}
\ee
This inequality can be proven by rewriting it in terms of the variable
$x\equiv D/(-V_t')$ as $4\geq x(x+1)^2$, and noting that $0\leq x\leq 1$ (as $V_t,V_t'\leq 0$ and $D$ should be real for $V_t$ to be an allowed decay path in the presence of gravity). To complete the proof, call $V_{t\kappa}$ and $V_{t0}$ the tunneling potentials that minimize the actions with and without gravity, respectively. Then we have
\be
S[V_{t\kappa}] \equiv \int_{\phi_\pp}^{\phi_0}{\it s}(V_{t\kappa}) d\phi\geq \int_{\phi_\pp}^{\phi_0}{\it s}_0(V_{t\kappa})d\phi\equiv S_0[V_{t\kappa}]\geq S_0[V_{t0}]\ ,
\ee
where the first inequality follows from (\ref{skgeqs0}) and the second from the fact that the tunneling functional $S_0[V_t]$ is minimized by $V_{t0}$.

\begin{figure}[t!]
\begin{center}
\includegraphics[width=0.6\textwidth]{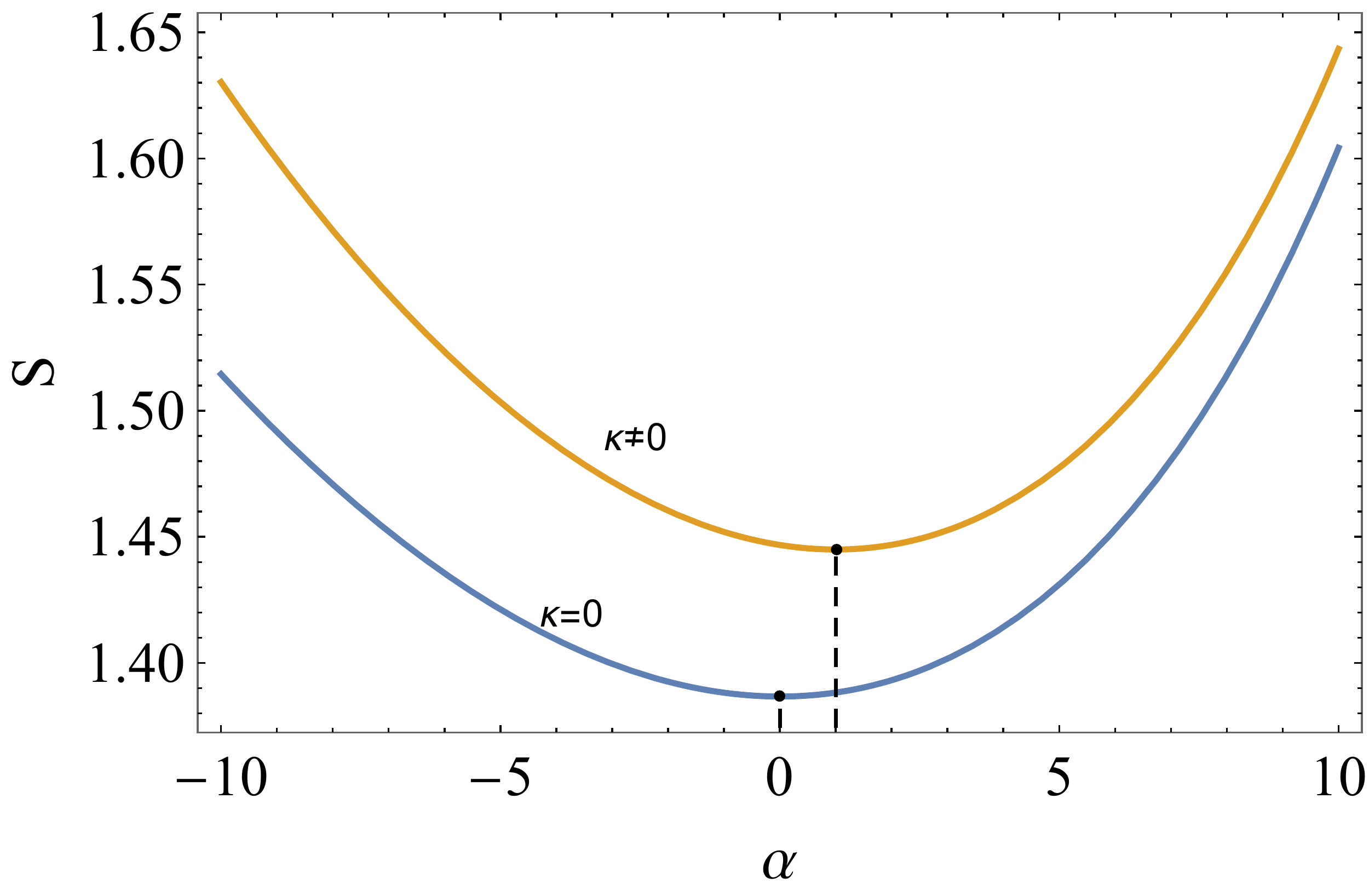}
\end{center}
\caption{Tunneling actions for a given potential with ($\kappa\neq 0$) and without gravity ($\kappa=0$) along a particular path (\ref{pathVt}) in $V_t$ space parametrized by 
$\alpha$. For this example, the analytic potential obtained from $V_{t0}(\phi)=\phi^2(2\phi-3\phi_\mm)$ in \cite{E} has been used, setting $\kappa\phi_\mm^2=1$.
\label{fig:comp}
}
\end{figure}

To illustrate the previous proof, Fig.~\ref{fig:comp} shows for a given potential the tunneling actions (with and without gravity, as indicated) along a particular path in $V_t$ space, parametrized by $\alpha$, with
\be
V_{t}(\alpha,\phi)\equiv (1-\alpha)V_{t0}(\phi)+\alpha V_{t\kappa}(\phi)\ ,
\label{pathVt}
\ee 
so that the path goes through the $V_t$ configurations that minimize both $\kappa=0$ and $\kappa\neq 0$ actions.
The corresponding action minima, at $\alpha=0,1$ are marked by  black dots. As the curve for $\kappa\neq 0$ is higher than the $\kappa=0$ at every point, the same inequality applies to the minima.

A different kind of proof of the same fact can be given  in the following way. Taking $\kappa$ as a variable, show that $dS[V_t]/d\kappa\geq 0$, so that the tunneling action is a monotonically increasing function of $\kappa$. 
This is also straightforward. One has
\be
\frac{dS[V_t]}{d\kappa}=\frac{\delta S}{\delta V_t}\frac{dV_t}{d\kappa}+ {\it s}(\phi_0)\frac{d\phi_0}{d\kappa}+\int_{\phi_\pp}^{\phi_0}\frac{\partial s(\phi)}{\partial \kappa}d\phi\ .
\ee
The first two terms vanish by using the equation of motion for $V_t$ and the boundary conditions and the remaining piece gives
\be
\frac{dS[V_t]}{d\kappa}=\frac{3\pi^2}{\kappa^3}\int_{\phi_\pp}^{\phi_0}\frac{(-V_t')}{V_t^2}\left(\frac{(-V_t')}{D}-1\right)^3\left(1+3\frac{D}{(-V_t')}\right)\, d\phi\geq 0\ ,
\ee
with the inequality following from $V_t'\leq 0$ and $0\leq D/(-V_t')\leq 1$.  The inequality holds order by order in the perturbative expansion, as shown explicitly at ${\cal O}(\kappa)$ by Eq.~(\ref{sexp}).

The line of argument in the first proof above can be used in a similar way to prove other inequalities, for instance, that higher barriers increase the tunneling action. If $V_{1,2}(\phi)$ are two potentials with the same minima at $\phi_\pp$ (metastable vacuum) and $\phi_\mm$ (deeper vacuum) and satisfying $V_2(\phi)\geq V_1(\phi)$, then 
\be
{\it s}_{2}(V_t)\equiv \frac{6\pi^2}{\kappa^2}\frac{(D_2+V_t')^2}{D_2 V_t^2}
\geq 
{\it s}_{1}(V_t)\equiv \frac{6\pi^2}{\kappa^2}\frac{(D_1+V_t')^2}{D_1 V_t^2}
\ ,\label{sk2geqsk1}
\ee
where $D_i^2=V_t'^2+6\kappa(V_i-V_t)V_t$ and we have $D_2\leq D_1$ (as $V_t\leq 0$ and $V_2\geq V_1$). Using the variables
$x_i\equiv D_i/(-V_t')$, that satisfy $0\leq x_i\leq 1$, and $x_2\leq x_1$, the inequality (\ref{sk2geqsk1}) takes the form $x_2+1/x_2\geq x_1+1/x_1$. This is satisfied for $x_2\leq x_1$ as $f(x)=x+1/x$ is monotonically decreasing in $(0,1)$. To complete the proof, call
$V_{t\kappa,i}$ the tunneling potentials corresponding to the potentials $V_i$ ({\it i.e.} giving the minimum of the respective actions). They are defined in different intervals $(\phi_\pp,\phi_{0,i})$, with $\phi_{0,i}\leq \phi_\mm$ being the exit points of the tunneling. The tunneling potential $V_{t\kappa,2}$ intersects the lower potential $V_1$ at some other field value $\phi_{0,21}\leq \phi_{0,2}$.
 Then we have
\bea
S_{2}[V_{t\kappa 2}]&\equiv &
\int_{\phi_\pp}^{\phi_{0,2}}{\it s}_{2}(V_{t\kappa 2})d\phi\geq 
\int_{\phi_\pp}^{\phi_{0,21}}{\it s}_{2}(V_{t\kappa 2})d\phi
\nonumber\\
&\geq & 
\int_{\phi_\pp}^{\phi_{0,21}}{\it s}_{ 1}(V_{t\kappa 2})d\phi
\geq 
\int_{\phi_\pp}^{\phi_{0,1}}{\it s}_{ 1}(V_{t\kappa 1})d\phi\equiv S_{1}[V_{t\kappa,1}] \ .
\eea
The first inequality follows from $\phi_{0,2}\geq\phi_{0,21}$ and the positivity of the action density; the second from (\ref{sk2geqsk1}) and the third from the fact that $V_{t\kappa 1}$ minimizes the action for $V_1$.
Notice that the argument does not require the inequality to hold for the action densities, ${\it s}_{\kappa 2}(V_{t\kappa 2})\geq {\it s}_{\kappa 1}(V_{t\kappa 1})$, which can be violated
(both in the tunneling potential and the Euclidean approaches).

\begin{figure}[t!]
\begin{center}
\includegraphics[width=0.42\textwidth]{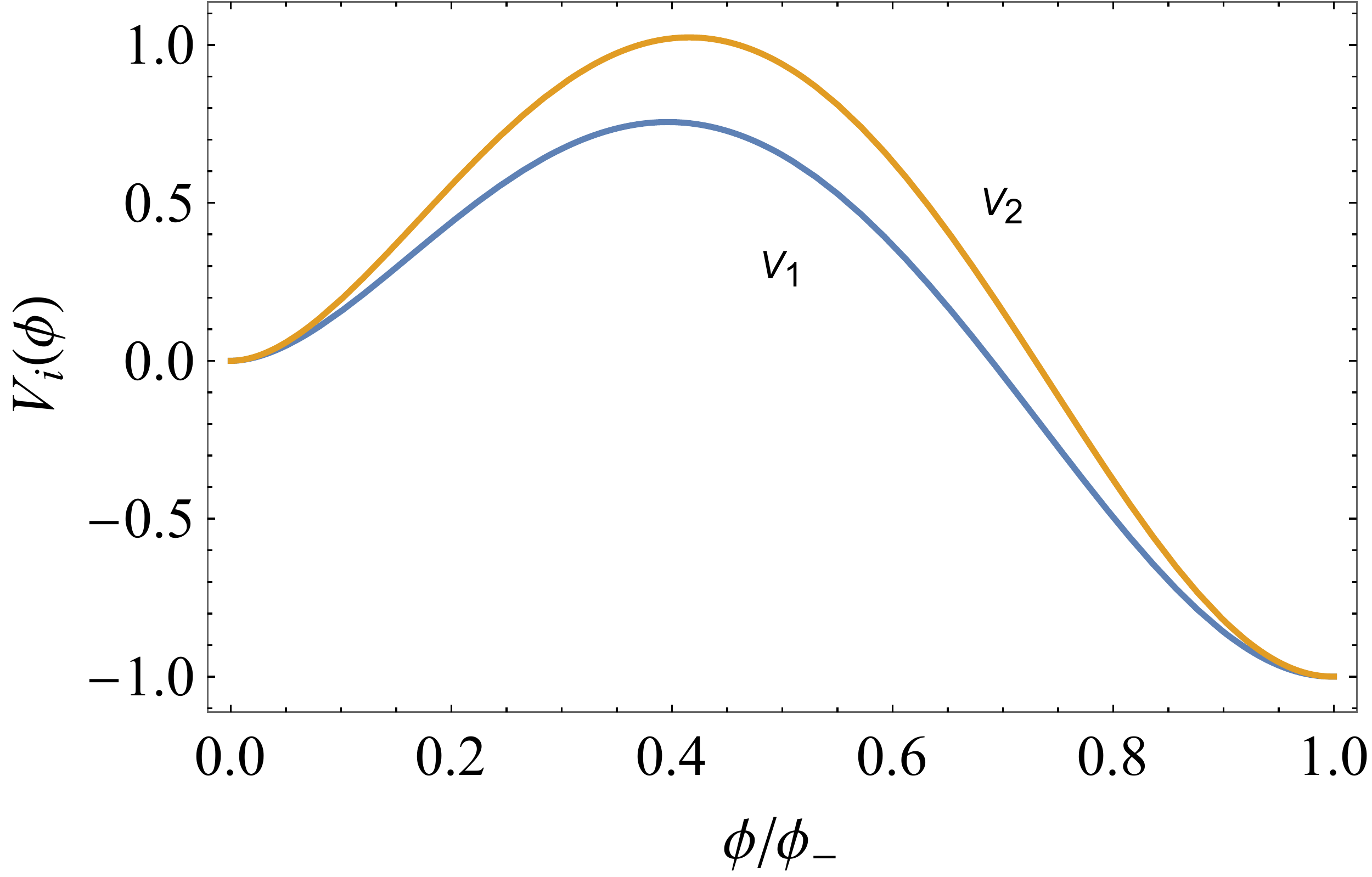}
\includegraphics[width=0.4\textwidth]{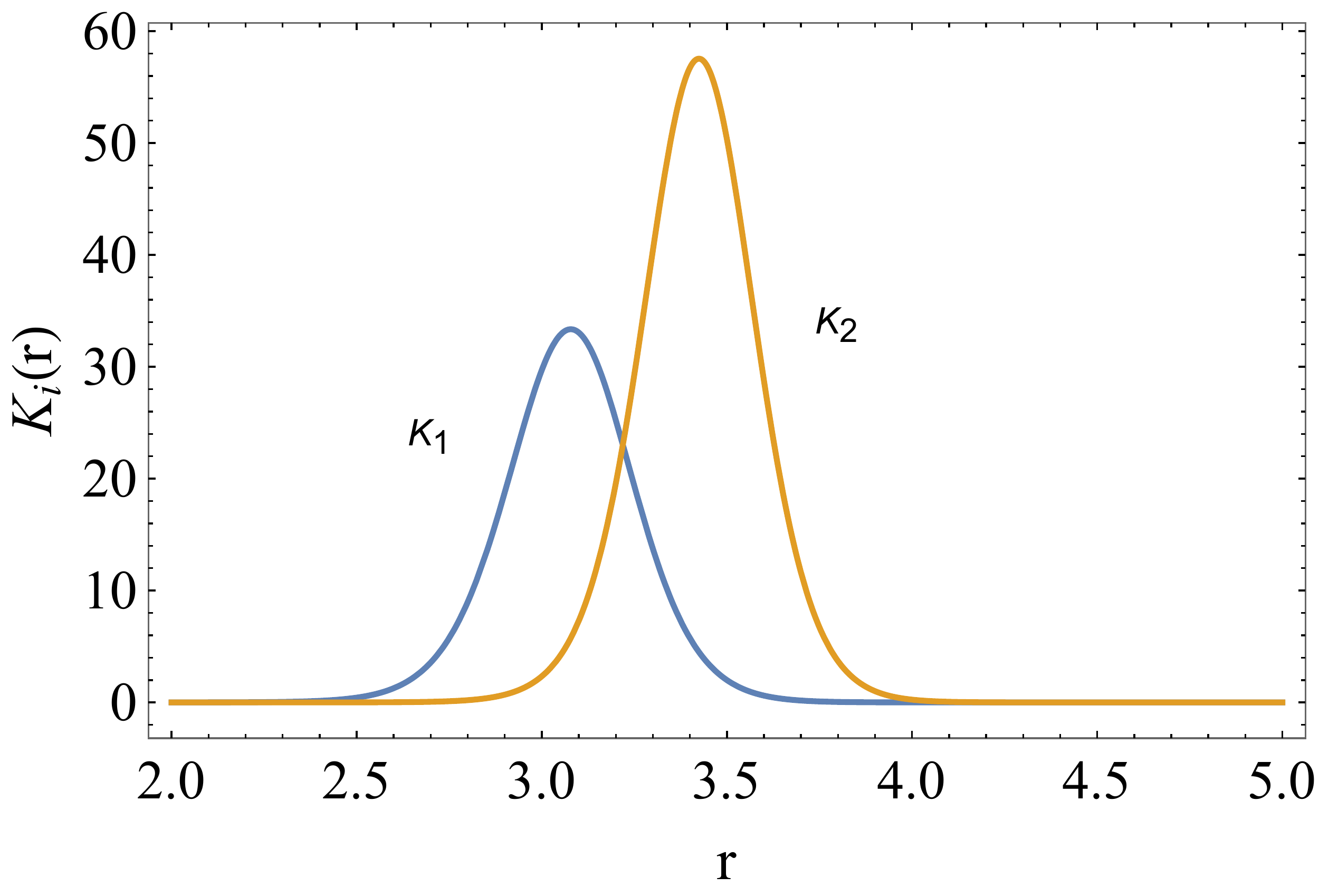}
\end{center}
\caption{Left Plot: Two potentials with the same minima and $V_2\geq V_1$. Right Plot: gradient contribution to the tunneling action, $K_i(r)\equiv 2\pi^2 r^3(d\phi_{B,i}/dr)^2/2$, where $\phi_{B,i}(r)$ is the Euclidean bounce for potential $V_i$ on the left. The action is $S_i=\frac12 \int_0^\infty K_i(r) dr$.
\label{fig:V1V2}
}
\end{figure}

As an illustration of this, Figure~\ref{fig:V1V2} shows two potentials with minima located at the same field values but with $V_2\geq V_1$ (left plot). For each potential, using the Euclidean bounce approach and Derrick's theorem \cite{Derrick}, one can calculate the corresponding tunneling action from the positive definite gradient contribution to the action as $S_i=\frac12 \int_0^\infty K_i(r) dr$. Here $K_i(r)\equiv 2\pi^2 r^3(d\phi_{B,i}/dr)^2/2$, where $\phi_{B,i}(r)$ is the Euclidean bounce for potential $V_i$. The right plot of Fig.~\ref{fig:V1V2} shows the $K_i(r)$ profiles, which do not
satisfy $K_2(r)\geq K_1(r)$.

\section{Gravitational Quenching\label{sec:quench}}

As known from the seminal paper of Coleman and De Luccia \cite{CdL} gravitational effects can completely stabilize metastable vacua leading to an infinite tunneling action that forbids vacuum decay (gravitational quenching). 
With the tunneling potential formulation we can easily estimate when gravity quenches vacuum decay. To have an allowed vacuum decay one needs to satisfy the condition 
\be
D^2 = V_t'{}^2+6\kappa (V-V_t)V_t>0\ ,
\label{quench}
\ee
and gravitational quenching of the decay happens when this condition cannot be satisfied no matter how $V_t$ is chosen \cite{Eg}. To see the impact of such strong gravitational effects we can consider making $\kappa$ larger (in the understanding that what is made larger is the dimensionless combination $\kappa M^2$, where $M$ is some characteristic mass scale of the potential). For Minkowski or AdS vacua the second term in (\ref{quench}) is negative and it is clear that for large enough $\kappa$ it will be impossible to satisfy the condition
(\ref{quench}) for any $V_t$ and the potential will be stabilized. This explains in a straightforward way previous results like those in \cite{EFT,MPW} that went beyond the thin-wall original analysis of Coleman and De Luccia \cite{CdL}.

We can estimate roughly when gravitational quenching would happen as follows. The order of magnitude of the tunneling potential roughly satisfies
$-V_t\sim (V_\pp+V_\mm)/2$, where $V_\pms\equiv V(\phi_\pms)$ ; then $-V_t'\sim \Delta V/\Delta\phi$, where $\Delta V\equiv V_\pp-V_\mm$ gives the energy difference between the two vacua and
$\Delta\phi\sim \phi_\mm-\phi_\pp$ is the field displacement in the tunneling process; finally, $V-V_t\sim \Delta V_T=V_T-V_\pp$, where $\Delta V_T$ measures the height of the barrier that separates the vacua.
Using these estimates on the quenching condition $D^2<0$ we get
\be
\boxed{
\frac{\Delta V}{\sqrt{3(\Delta V-2V_\pp)\Delta V_T}}\simlt  \frac{\Delta\phi}{m_P}}\ ,
\label{quench0}
\ee
as the rough condition to have gravitational quenching of vacuum decay. This formula encapsulates the expected parametric behaviour needed for quenched potentials: large enough $\Delta\phi$, high potential barriers (that make $\Delta V_T$ large), very shallow true minima (giving small $\Delta V$) or very deep AdS false minima (making $-V_\pp$ large). Such behaviour was already discussed numerically or semi-analytically in \cite{EFT} (see also \cite{MPW}) but the tunneling potential approach allows for a very direct and simple understanding of this quenching effect.

We can go beyond the order of magnitude estimate (\ref{quench0})
for gravitational quenching and analyze in quantitative detail how it comes about. We can interpret the condition $D^2>0$ to have vacuum decay as meaning the following: with gravity, for AdS or Minkowski vacua, $V_t$ has to satisfy a condition stronger than mere monotonicity:
\be
V_t'\leq -\sqrt{6\kappa (V-V_t)(-V_t)}\ ,
\label{newmono}
\ee
to have $D$ real. Notice that this gives back the monotonicity condition \cite{E} in the absence of gravity ($\kappa\rightarrow 0$).

For a given potential $V$ and a given metastable minimum at $\phi_\pp$ we can obtain what we will call the critical tunneling potential, $V_{tc}$, as the solution to  $D=0$ with
\be
V_{tc}'= -\sqrt{6\kappa (V-V_{tc})(-V_{tc})}\ ,
\label{Vtc}
\ee
with boundary condition $V_{tc}(\phi_\pp)=V(\phi_\pp)\equiv V_\pp$.
It is important to note that, in order to integrate (\ref{Vtc}) and obtain the critical $V_{tc}$, it is sufficient to have the boundary condition at $\phi_\pp$ (unlike what happens with the tunneling potential for vacuum decay, which requires to satisfy the proper boundary conditions both at $\phi_\pp$ and $\phi_0$).
Other solutions of (\ref{Vtc}) with different boundary values for $V_{tc}(\phi_\pp)$ generate a family of non-intersecting integral curves for $D=0$ that cover 
the area below $\mathrm{Min}\{V_\pp,V\}$.

\begin{figure}[t!]
\begin{center}
\includegraphics[width=0.45\textwidth]{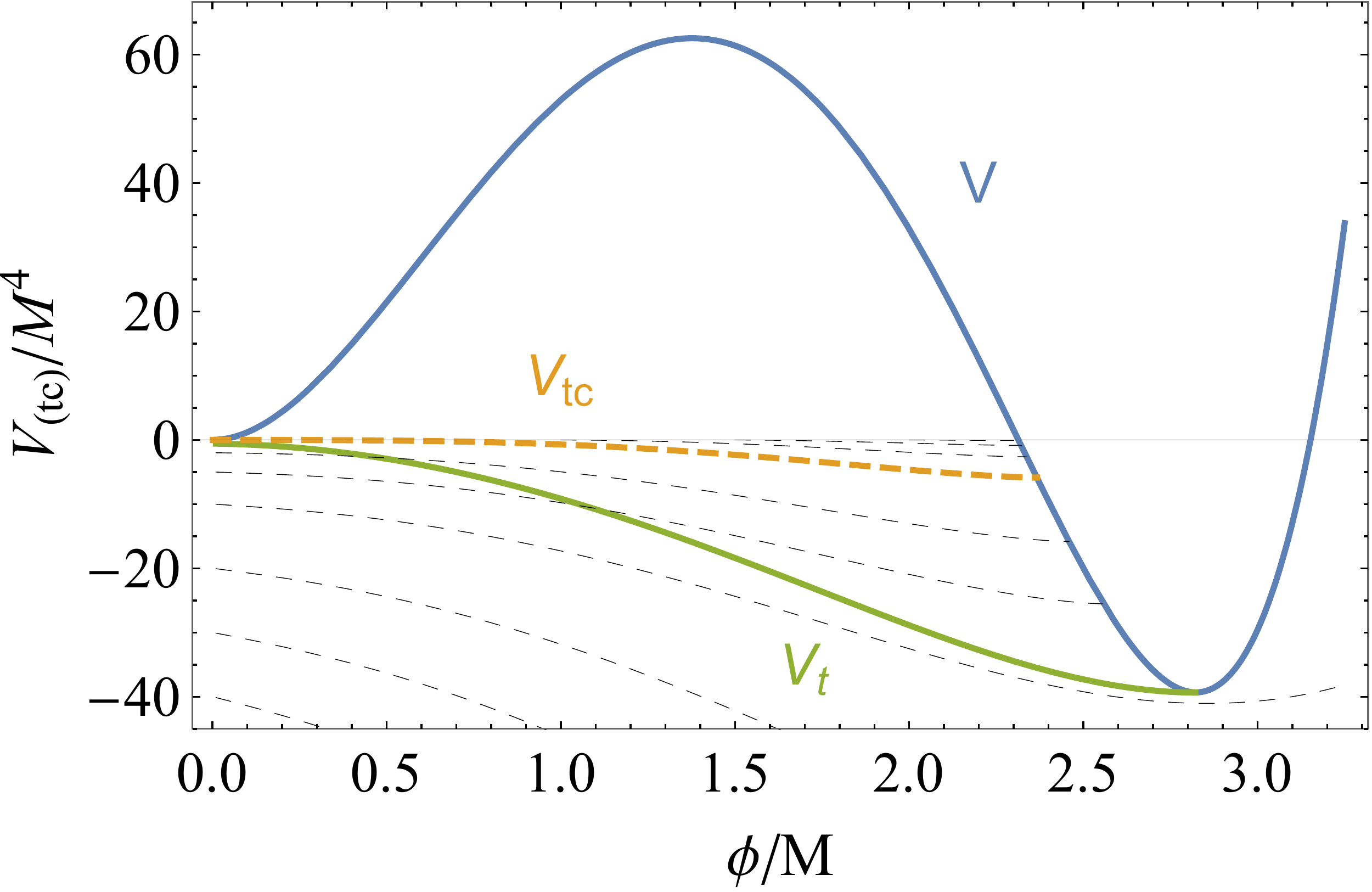}
\includegraphics[width=0.45\textwidth]{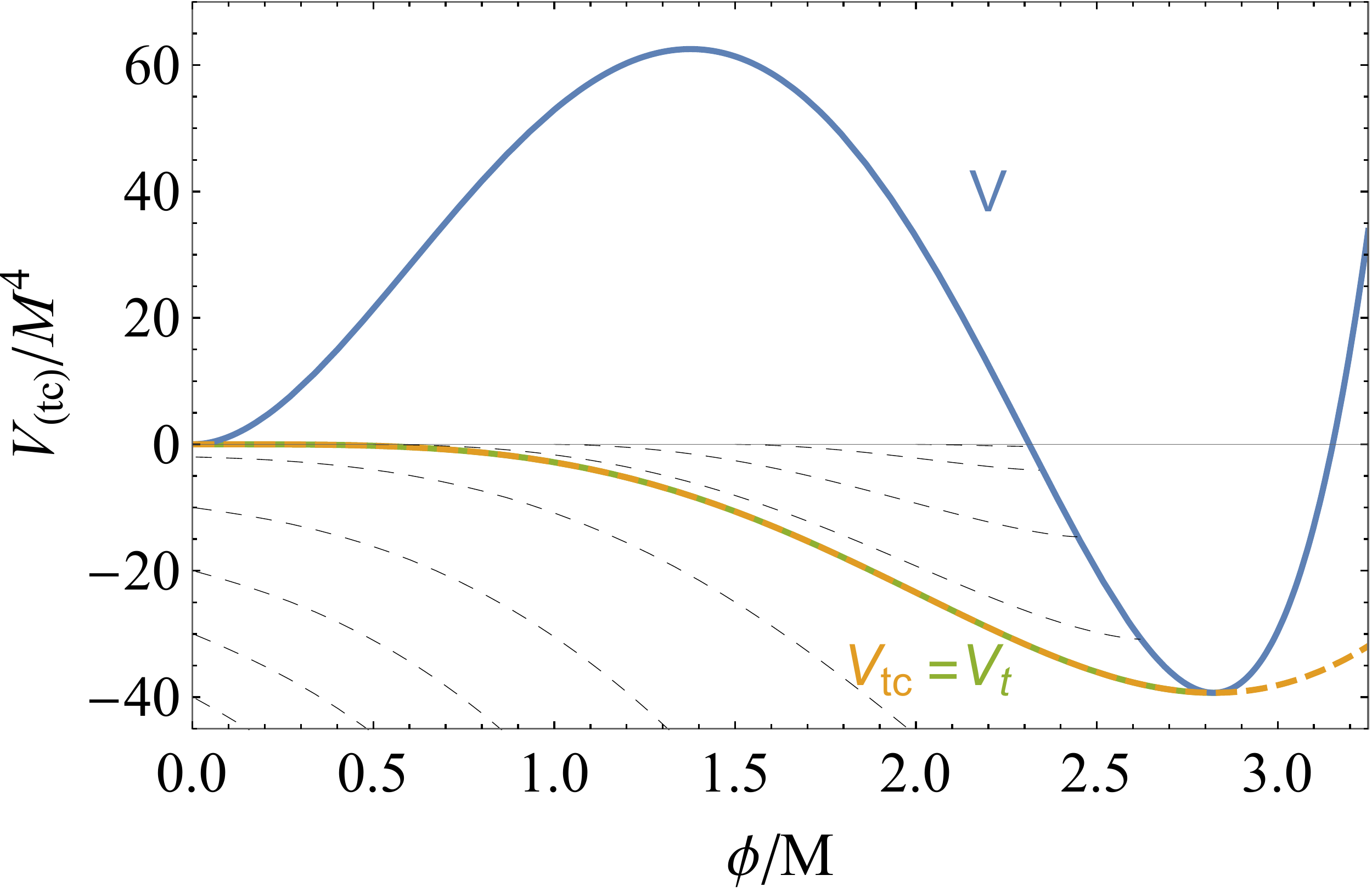}\\
\vspace*{0.5cm}
\includegraphics[width=0.5\textwidth]{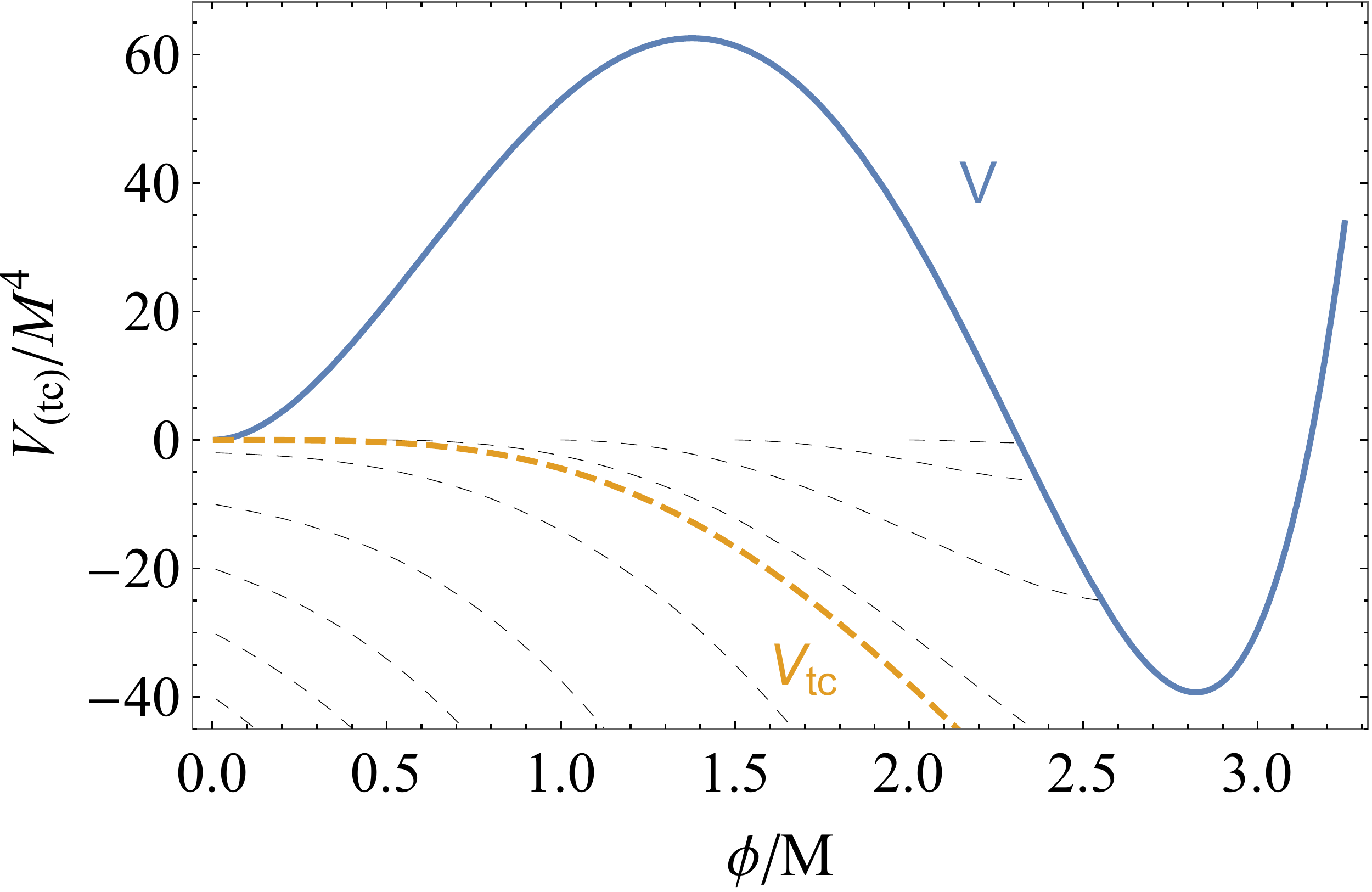}
\end{center}
\caption{Potential with a Minkowski false vacuum at $\phi_+=0$  for different strengths of gravitational effects as measured by $\kappa M^2$. Upper left plot: subcritical case, with $\kappa M^2=0.02$; Upper right plot: critical case with $\kappa M^2=0.1$; Lower plot: supercritical case with $\kappa M^2=0.15$. In all cases, dashed lines are integral curves of $D=0$ with $V_t'\leq 0$ and the orange dashed line is the critical tunneling potential satisfying (\ref{Vtc}) and $V_{tc}(\phi_+)=0$. In the first two cases the tunneling potential for vacuum decay is given by the green line. \label{fig:kV}
}
\end{figure}

We illustrate this in Fig.~\ref{fig:kV}, which shows, for a given potential shape with different values of $\kappa M^2$, such family of integral curves by dashed lines, with $V_{tc}$ singled out as indicated. According to (\ref{newmono}), the tunneling potential $V_t$ describing the decay of $\phi_\pp$  can only intersect these integral lines from above and must lie below $V_{tc}$. For comparison, the $\kappa=0$ case is trivial: the integral lines are simply horizontal lines and $V_{tc}$ is the horizontal line $\phi=\phi_\pp$. Let us consider three different cases, depending on the strength of gravitational effects.

{\bf Subcritical case.} Fig.~\ref{fig:kV}, upper left plot, shows the typical case with weak gravitational effects: the critical $V_{tc}$ deviates somewhat from being horizontal but reaches $V$ far from $\phi_\mm$ and leaves ample space for $V_t$ to lie below it, intersect the $D=0$ integral lines from above and reach $V$ at some $\phi_0$. In such (subcritical) cases, gravity makes the $\phi_\pp$ vacuum slightly more stable but does not forbid its decay.  

{\bf Critical case.} Fig.~\ref{fig:kV}, upper right plot, corresponds to the critical case
for which $V_t\equiv V_{tc}$. Notice that $D=0$ solves the EoM for the tunneling potential (\ref{EoMVtD}) so, the critical case shown in the figure corresponds to $V_{tc}$ satisfying the right boundary condition (\ref{BCsVtp}) at $\phi_0=\phi_\mm$. For this critical case, the tunneling action is infinite, and gravity forbids the decay of $\phi_\pp$ into $\phi_\mm$. This critical case for Minkowski vacua corresponds to the so-called ``great divide'' case of \cite{GDivide} (approached in that paper  from the de Sitter side, with $V_\pp\rightarrow 0^+$, see also \cite{BFL}).

The fact that both vacua are connected by $V_{tc}$ plays the same role as having degenerate minima in the absence of gravity: no decay is allowed, although one can have a static domain wall separating the two vacua (even though  they have different scalar potential vacuum energies). Such static domain wall is the infinite radius limit of a Coleman-De Luccia (CdL)  bubble and its tension can be computed in closed form as
\be
\sigma\equiv\int_{\phi_\pp}^{\phi_\mm}\sqrt{2(V-V_t)}d\phi=-\int_{\phi_\pp}^{\phi_\mm}\frac{V_t'}{\sqrt{-3\kappa V_t}}d\phi=2\left.\sqrt{\frac{-V_t(\phi)}{3\kappa}}\right|^{\phi_\mm}_{\phi_\pp}=2\left.\sqrt{\frac{-V(\phi)}{3\kappa}}\right|^{\phi_\mm}_{\phi_\pp}\ ,
\label{wtdw}
\ee
where the second step follows from $D=0$ (with $V_t'\leq 0$). 

If we solve the condition $D\equiv 0$ for $V$ we obtain the result that any potential made critical by gravity (regarding vacuum decay) must take the generic form
\be
\boxed{
V_c(\phi)=V_t-\frac{V_t'{}^2}{6\kappa V_t}}\ ,
\label{Vc}
\ee
for some monotonic function $V_t(\phi)$. This reproduces in a straightforward way the old results of \cite{Boucher,AP}.
There is a large class of theories that have such critical vacua: supersymmetric ones. That particularly important case is discussed in more detail in Section~\ref{sec:SUSY}.

{\bf Supercritical case.} Finally, the lower plot of Fig.~\ref{fig:kV} shows a case of strong gravitational effects curving down the $V_{tc}$ line so much that it does not intersect $V$ (for $\phi>\phi_\pp$). As $V_t$ must lie below $V_{tc}$, this forbids the existence of a viable $V_t$ with real $D$: vacuum decay is forbidden by gravity. 
It seems tempting to conclude from this that the opposite decay, from $\phi_\mm$ towards $\phi_\pp$ should then be possible, but this is not the case: $V_t$ should decrease monotonically out of an AdS (or Minkowski) decaying vacuum \cite{Eg} and that is  impossible for $V_\pp>V_\mm$.

In summary, the recipe to find out if a given Minkowski or AdS vacuum is allowed to decay is to solve (\ref{Vtc}) with $V_{tc}(\phi_\pp)=V_\pp$ and check whether it does intersect $V$ or not. The critical case $V_t=V_{tc}$ corresponds to the limiting case of an intersection precisely at the minimum $\phi_\mm$.

\section{Stability of Supersymmetric Vacua\label{sec:SUSY}}

The gravitational stabilization of supersymmetric vacua \cite{SWeinberg,Cvetic} is of particular interest. Ultimately, the reason why supersymmetric vacua do not decay into each other can be related \cite{SUSYvacua} to general theorems of the positivity of energy \cite{Positivity} (which forbid the existence of the 
zero-energy  bubbles mediating vacuum decay). Here we analyze this particular case using the same tunneling potential method of previous sections.

The scalar potential in $N=1$ supergravity reads
\be
V(\Phi^i)=e^{\kappa K}\left[D_iW\,\overline{D_jW}g^{i\overline j}-3\kappa|W|^2\right]\ ,
\label{Vsusy}
\ee
where $K$ is the K\"ahler potential (a real function of the complex fields $\Phi^i$), $W$ is the superpotential (a holomorphic function of the $\Phi^i$'s), $g^{i\overline j}$ is the inverse of the (positive definite) K\"ahler metric 
$g_{\overline i j} \equiv \partial^2K/\partial\Phi^i{}^*\partial\Phi^j$, $D_iW\equiv \partial W/\partial\Phi^i+\kappa W \partial K/\partial\Phi^i $ and an overline indicates complex conjugation. Supersymmetric vacua satisfy $D_iW=0, \forall i$ and can be either Minkowski or AdS, depending on the value of $|W|$. In particular, $V=0$ is a natural value for a supersymmetric vacuum.

Consider a potential of the form (\ref{Vsusy}) with two supersymmetric minima: a false one at $\Phi^i_\pp$, where $V_\pp=-3\kappa e^{\kappa K_\pp}|W_\pp|^2$ and a deeper one at $\Phi^i_\mm$, where $V_\mm=-3\kappa e^{\kappa K_\mm}|W_\mm|^2$, with $V_\mm<V_\pp$ (the subindices $_\pms$ indicate that the function is evaluated at $\Phi^i=\Phi_\pms^i$). We want to address the possible decay out of $\Phi_\pp$ towards $\Phi_\mm$. 
The potential (\ref{Vsusy}) is a multifield one but the tunneling potential approach can be easily extended to such cases \cite{EK}. 
To find the tunneling action for a multifield potential $V$  one has to find the trajectory in field space out of the metastable vacuum and the tunneling potential along such trajectory so as to minimize the single-field action (\ref{SVt}). In that action, field derivatives are taken with respect to a field $\phi$ that canonically parametrizes the trajectory $\Phi^i(\phi)$ between the vacua with
\be
g_{\overline i j}\ d\Phi^i{}^* d\Phi^j=\frac12 d\phi^2\ .
\ee
That is, $\phi$ is the arc length along that trajectory and we take $\Phi^i(\phi_\pp=0)=\Phi^i_\pp$.
Notice that this single-field reduction works even if the kinetic term
in the multifield case is not canonical, that is, $g_{\overline i j}(\Phi^k)\neq \delta_{ij}$. More details about the multifield case, including the particular case of complex fields as needed in supersymmetry can be found in Appendix \ref{sec:app}.

It is straightforward to check that the tunneling potential is simply\footnote{In terms of the gravitino mass, we would write
$V_t=-m_p^2\, m_{3/2}^2(\Phi^i,\Phi^{i*})$.}
\be
V_t=-3\kappa|W|^2e^{\kappa K}\ ,
\label{Vtsusy}
\ee
taken along a field direction that satisfies\footnote{For the field trajectories of interest to us (limiting cases of CdL bubbles) $W$ or $DW$ do not vanish at intermediate points between two vacua and (\ref{SUSYpath}) is well defined. An example of alternative trajectory with $W= 0$ at some intermediate point  is  discussed later on in this section. The simplest such case is a double-well potential with two AdS degenerate minima \cite{Cvetic}.}
\be
\frac{d\Phi^i}{d\phi} = \frac{g^{i\overline j} W \overline{D_j W}}{\sqrt{2}|W||DW|}\ ,
\label{SUSYpath}
\ee
where $|DW|^2\equiv D_i W g^{i\overline j} \overline{D_j W}$. With these choices one indeed has $D=0$. Note that 
\be
\frac{d\Phi^i}{d\phi} \propto g^{i\overline j}\frac{\partial V_t}{\partial \Phi^j{}^*}\ ,
\ee
so that the tunneling path follows the gradient of $V_t$.
For the particular trajectory $\Phi^i(\phi)$ with end points corresponding to the two SUSY minima, with $V_{t\pms}=V_\pms$, one has $D=0$, so that the decay is forbidden, as in the critical non-supersymmetric cases discussed in the previous section. 

\begin{figure}[t!]
\begin{center}
\includegraphics[width=0.35\textwidth]{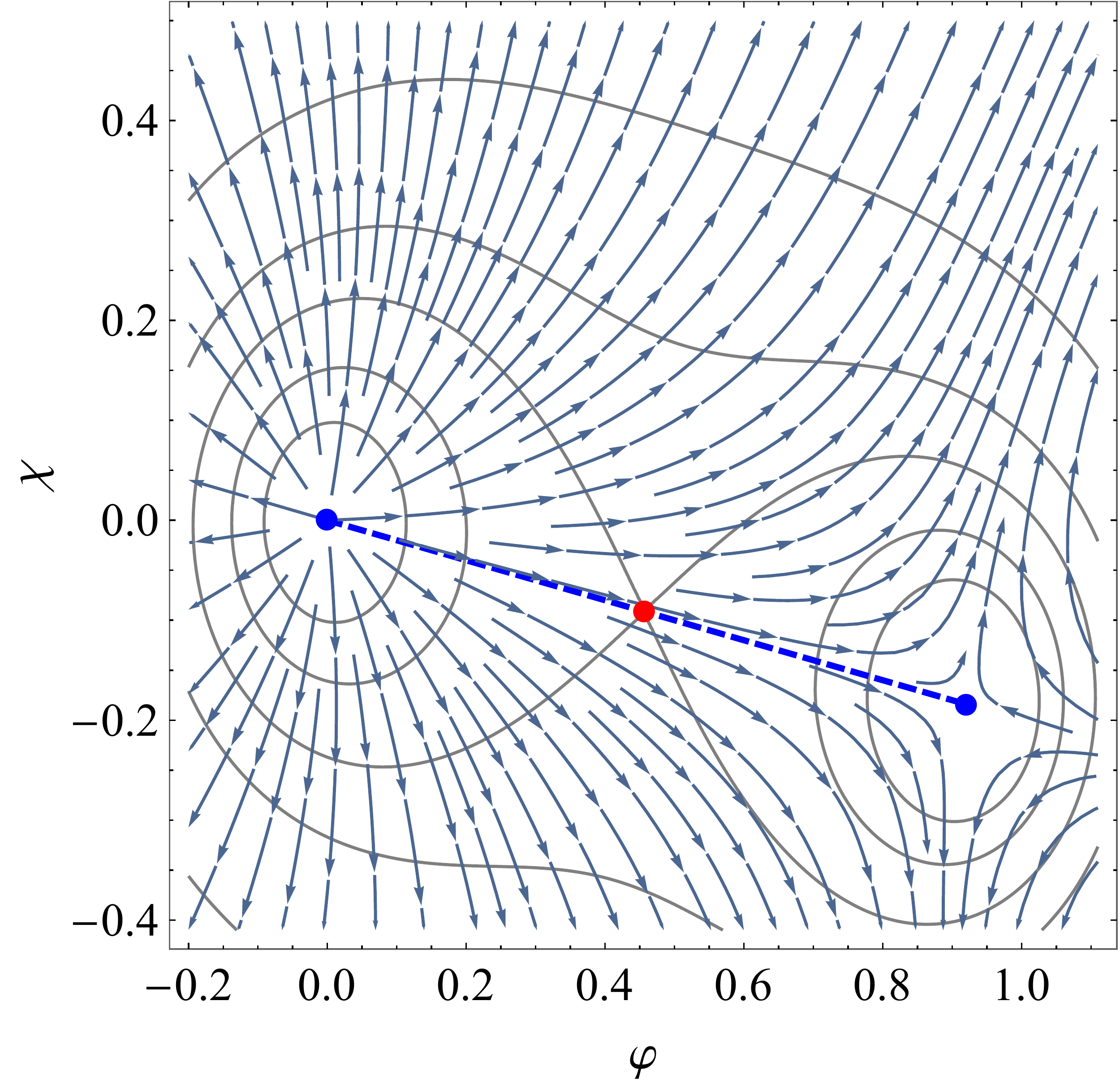}\hspace*{0.5cm}
\includegraphics[width=0.5\textwidth]{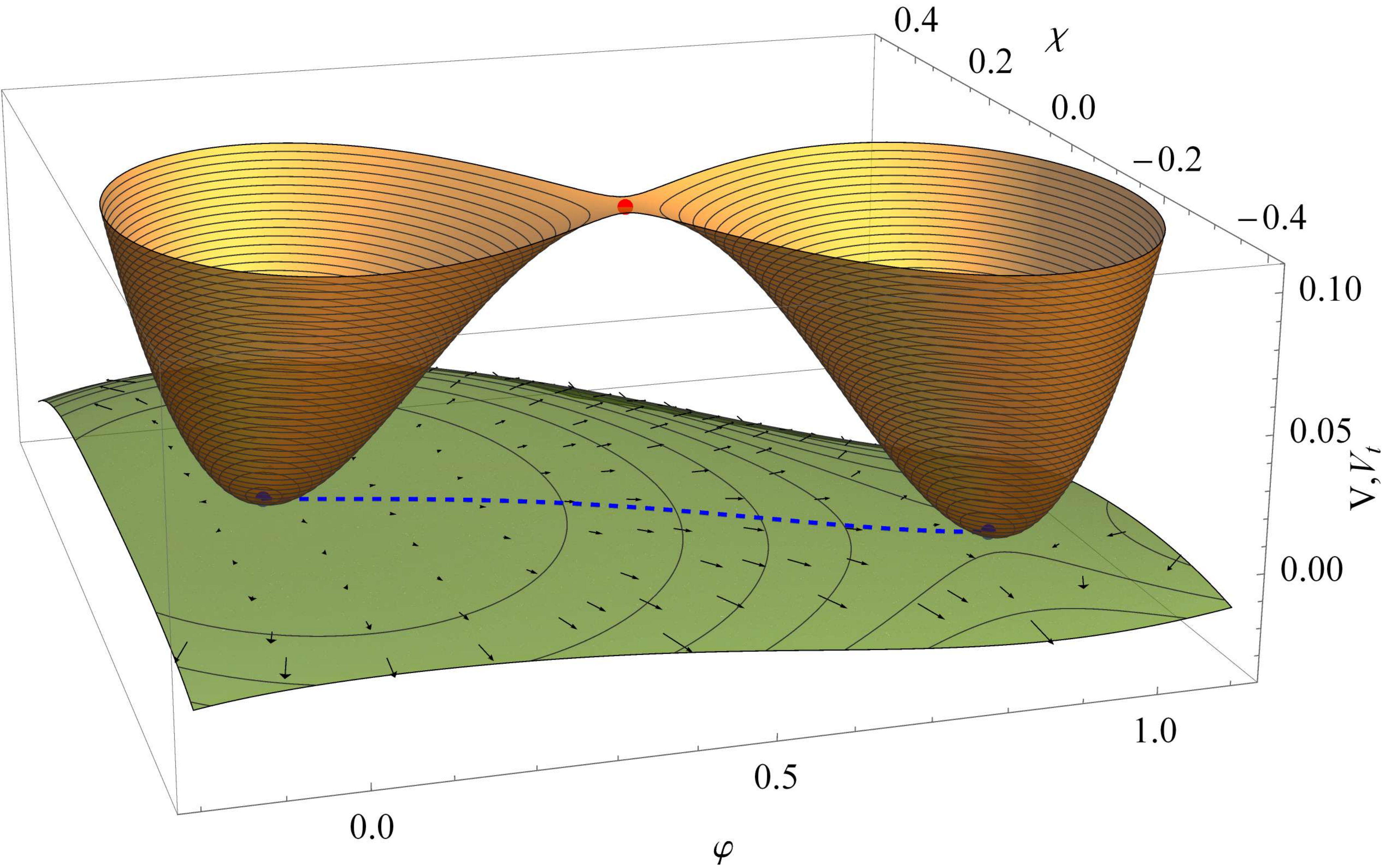}
\end{center}
\caption{For the potential (\ref{Vsusytosusy}) with
$m=2$, $\kappa=0.2$ (in arbitrary mass units) and $\lambda=3+0.6i$, the left plot gives contours of fixed $V$, with the minima marked by blue dots and the saddle point by a red dot. The streamlines correspond to (\ref{SUSYpath}) and the dashed blue line to the particular path $\Phi(\phi)$ joining the two minima. The right plot is the 3D version of the same and shows $V$, $V_t$ (covered by its gradient vector field) and the path $\Phi(\phi)$ projected on $V_t$.
\label{fig:VSUSYtoSUSY}
}
\end{figure}

Let us illustrate the previous discussion with a simple example.
Take the superpotential 
\be
W=\frac12 m\Phi^2-\frac13\lambda \Phi^3\ ,
\label{Wsusytosusy}
\ee
with $m$ real and positive and $\lambda$ complex,
and canonical K\"ahler potential $K=|\Phi|^2$. 
The potential is
\be
V= e^{\kappa|\Phi|^2}\left[|\Phi|^2\left|m-\lambda\Phi +\kappa |\Phi|^2 \left(\frac12 m-\frac{\lambda}{3}\Phi\right)\right|^2-3\kappa\left|\frac12 m\Phi^2-\frac{\lambda}{3}\Phi^3\right|^2\right]\ .
\label{Vsusytosusy}
\ee
Writing $\Phi=(\varphi +i \chi)/\sqrt{2}$, $V$ is a function of the two real fields $\varphi$ and $\chi$. 
This potential (\ref{Vsusytosusy}) has a false supersymmetric vacuum at $\varphi=\chi=0$ with $V_\pp=0$ and generically has at least another deeper supersymmetric vacuum at $\varphi,\chi\neq 0$, with $\varphi$ being the real root of the cubic polynomial  $m-|\lambda|^2x+m|\lambda|^2\kappa x^2/4-|\lambda|^4\kappa x^3/6$, with $x\equiv \varphi/(\sqrt{2}\mathrm{Re}\lambda)$
and $\chi/\varphi=-\mathrm{Im}\lambda/\mathrm{Re}\lambda$. 

Figure~\ref{fig:VSUSYtoSUSY} shows one particular example of potential (\ref{Vsusytosusy}) for the indicated choice of  parameters. The left plot gives contour lines of $V(\varphi,\chi)$, with two minima marked by blue dots (the one at the origin and a deeper one) and a saddle point (on the top of the barrier that separates both vacua) in red. The streamlines shown correspond to the vector field $(d\varphi/d\phi,d\chi/d\phi)$ as calculated from (\ref{SUSYpath}). The particular streamline joining the two minima is shown by the dashed blue line. The right 3D-plot of Fig.~\ref{fig:VSUSYtoSUSY} shows the potential and its minima and also $V_t$, which is tangent to $V$ at the minima. The vector field corresponds to $(d\varphi/d\phi,d\chi/d\phi)$  projected onto $V_t$ and is clearly aligned with the gradient of $V_t$. We also show the path along $V_t$ that joins the minima.

As was discussed in the previous section, minima joined by a path along which $D=0$ can support a static domain wall separating them. In the supersymmetric case, the wall tension for such BPS domain walls \cite{BPS} separating supersymmetric vacua can be obtained, plugging $V_t$ of (\ref{Vtsusy})
in (\ref{wtdw}), as
\be
\sigma=\left.2e^{\kappa K/2}|W|\right|_\pp^-\ ,
\ee
saturating the so-called Bogomol'nyi bound.

From the previous discussion and example, one should not conclude that every false supersymmetric vacuum is critical. One of the conditions $V_t$ should satisfy is monotonicity, with $dV_t/d\phi\leq 0$. However, it is not guaranteed that the supersymmetric expression for $V_t$ given in (\ref{Vtsusy}) is monotonically decreasing along (\ref{SUSYpath}) for all $W$ and $K$ (notice that $D\equiv 0$ fixes
$V_t'$ only up to an overall sign). Let us examine now a potential (a modification of the KKLT potential, taken from \cite{CDGKL}) that features a false supersymmetric vacuum that is supercritical. Consider the superpotential (we set $\kappa=1$ in this example for simplicity)
\be
W=W_0+A e^{-a \rho} + B e^{-b\rho}\ ,
\ee
where $\rho=e^{2\varphi/\sqrt{3}}+i \alpha$ is a volume modulus 
field, and the K\"ahler potential  $K=-3\log (\rho+\bar\rho)$. 
The potential is
\bea
V&=& \frac{1}{(\rho+\bar\rho)^3}\left\{\frac13 (\rho+\bar\rho)^2
\left|a A e^{-a \rho} + b B e^{-b\rho}+\frac{3}{\rho+\bar\rho}\left(W_0+A e^{-a \rho} + B e^{-b\rho}\right)\right|^2
\right.\nonumber\\
&-&\left.
3\left|W_0+A e^{-a \rho} + B e^{-b\rho}\right|^2
\right\} .
\label{VSUSYW0}
\eea
With the choice of parameters $A=1.05$, $B=-2$, $W_0=-0.125$, $a=\pi/100$ and $b=\pi/50$ as in \cite{CDGKL}, the potential $V(\varphi,a)$ has two AdS supersymmetric minima with  $\varphi_\pp=3.22$ and  $\varphi_\mm=3.97$ (both with $\alpha=0$) with $V_\mm<V_\pp<0$ (and asymptotes to Minkowski for $\varphi\rightarrow \infty$).

\begin{figure}[t!]
\begin{center}
\includegraphics[width=0.35\textwidth]{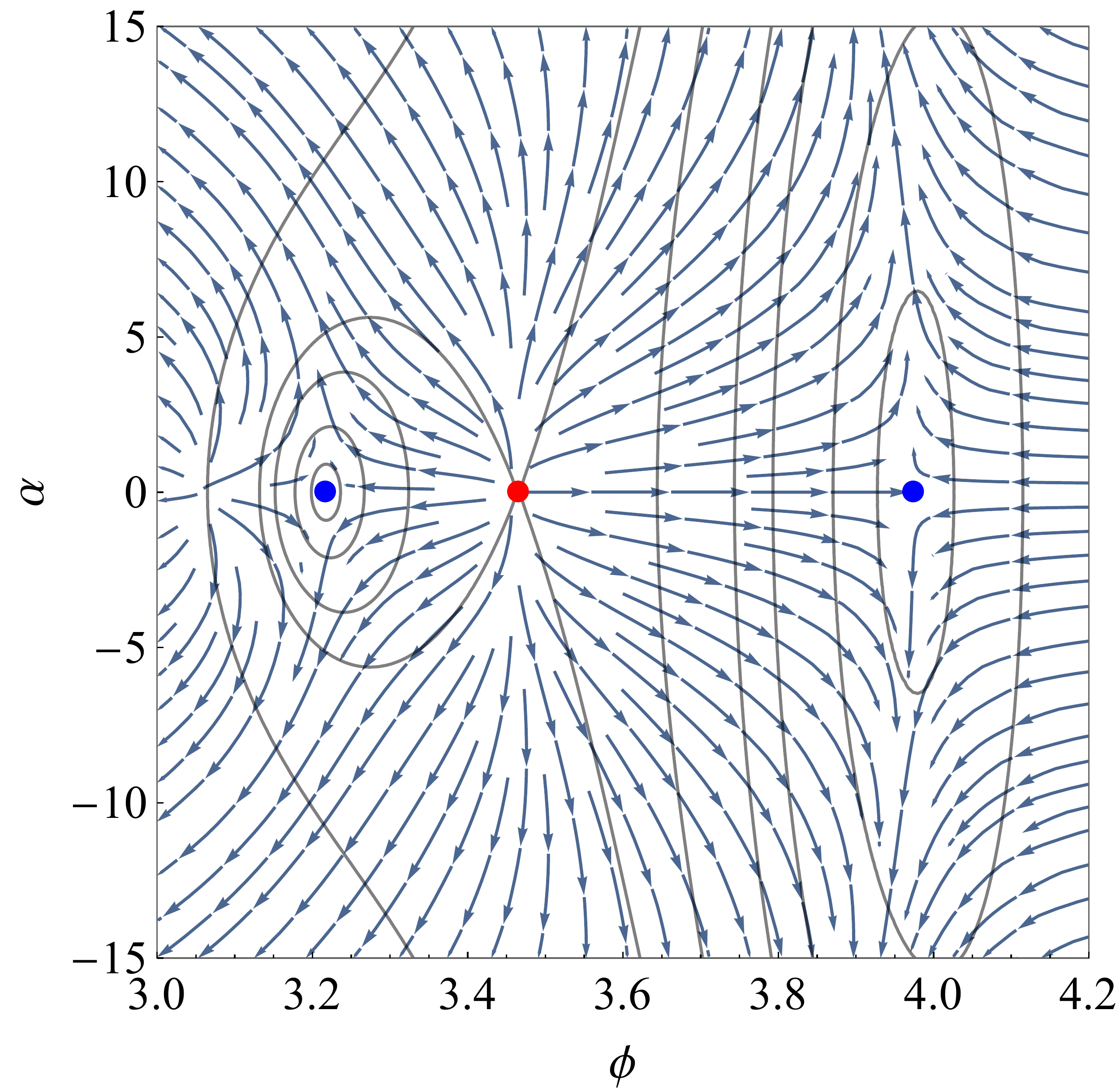}\hspace*{0.5cm}
\includegraphics[width=0.5\textwidth]{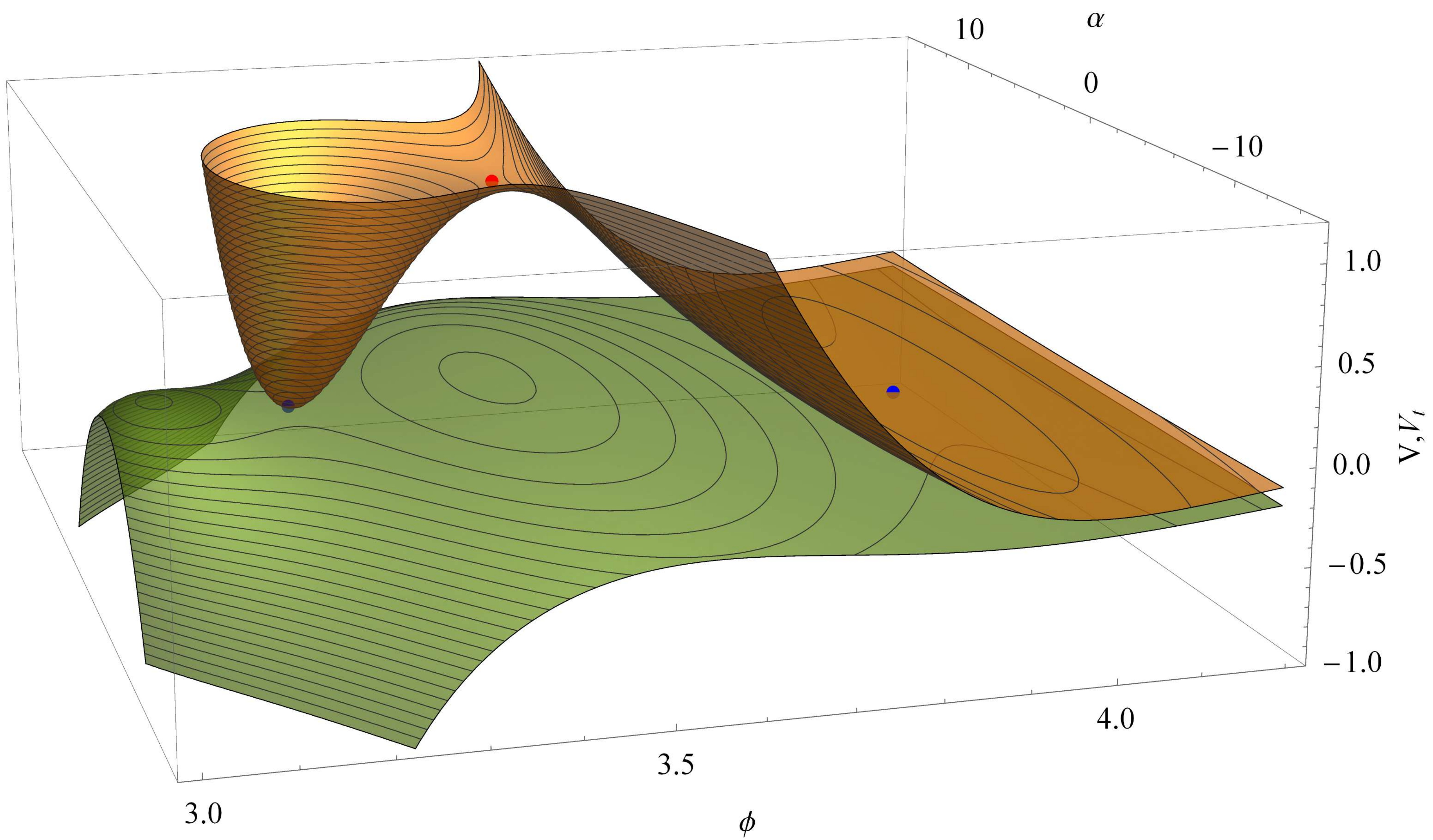}
\end{center}
\caption{For the potential (\ref{VSUSYW0}) with  $A=1.05$, $B=-2$, $W_0=-0.125$, $a=\pi/100$ and $b=\pi/50$, the left plot gives contours of fixed $V$, with the minima marked by blue dots and the saddle point by a red dot. The streamlines correspond to (\ref{SUSYpath}). The right plot is the 3D version of the same and shows $V$ and $V_t$. (In both plots, $V$ and $V_t$ are rescaled by $10^8$.)
\label{fig:VSUSYW0}
}
\end{figure}

\begin{figure}[t!]
\begin{center}
\includegraphics[width=0.5\textwidth]{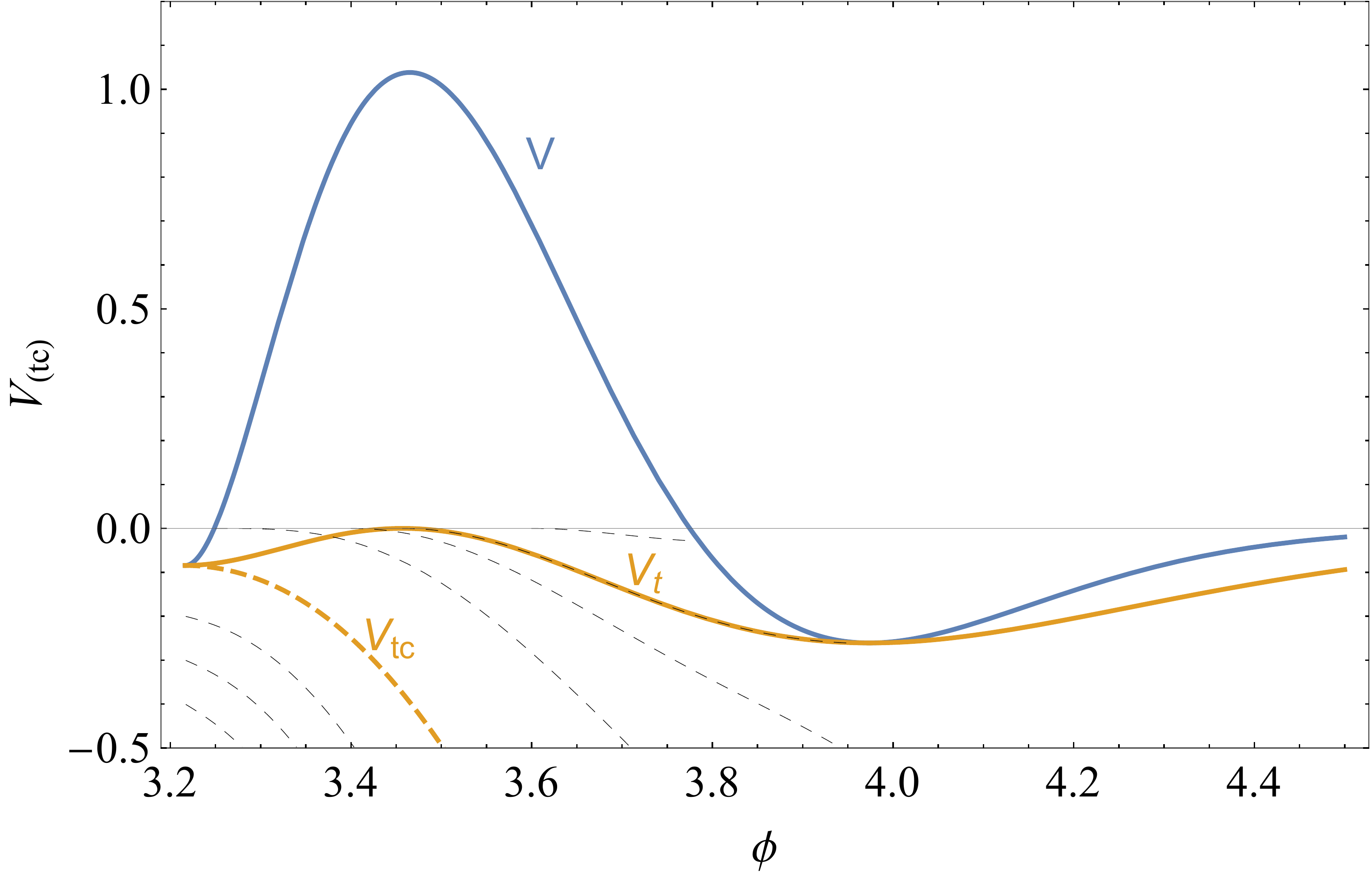}
\end{center}
\caption{Potential of Fig.~\ref{fig:VSUSYW0} for fixed $\alpha=0$  and corresponding $V_t$, both rescaled by a factor $10^8$. For the false vacuum the dashed lines give monotonically decreasing lines with $D=0$, with $V_{tc}$ singled out. The $V_t$ joining the minima corresponds to a supercritical domain wall.
\label{fig:VSUSYKL}
}
\end{figure}

Figure~\ref{fig:VSUSYW0} shows this potential for
the indicated choice of parameters. The left plot gives contour lines of $V(\varphi,\alpha)$ with the two AdS minima marked by blue dots (with the false one on the left) and the intermediate saddle point in red. The streamlines shown correspond to the vector field $(d\varphi/d\phi,d\alpha/d\phi)$ as dictated by (\ref{SUSYpath}) and follow the gradient of $V_t$. Lines leaving the false vacuum do not reach the true vacuum on the right: the saddle point of $V$ corresponds to a maximum of $V_t$ and this prevents the flow from false to true vacuum. The right 3D-plot of Fig.~\ref{fig:VSUSYW0} shows the potential and its minima and also $V_t$, which is tangent to $V$ at the minima but has a maximum in between. That maximum corresponds to a point with $W=0$.  Figure~\ref{fig:VSUSYKL} shows $V$ and $V_t$ along the line joining the minima, with $\alpha=0$. The figure also shows explicitly that the AdS false vacuum on the left is indeed supercritical: the dashed lines give the monotonically decreasing $D=0$ curves, with the one corresponding to the critical $V_{tc}$ highlighted. Notice also that the (monotonically decreasing) branch of $V_t$ on the right of the maximum follows one of the (dashed) $D=0$ lines, as it should. The (monotonically increasing) branch on the left of the maximum gives also $D=0$ but has the wrong sign, $V_t'\geq 0$. The complete $V_t$ from minimum to minimum corresponds in fact to a different\footnote{The domain wall tension in this case can also be derived as in (\ref{wtdw}) paying attention to the sign $\zeta$ of $V_t'$. One gets $\sigma=-2\zeta\left.\sqrt{-V(\phi)/(3\kappa)}\right|_\pp^-=2(e^{\kappa K_\mm/2}|W_\mm|+e^{\kappa K_\pp/2}|W_\pp|)$, larger than the Bogomol'nyi bound.} type of domain-wall \cite{Cvetic,CDGKL} not directly related to a limiting case of CdL bounce and is beyond the scope of this paper.  (The piece of $V_t$
that runs from the deeper minimum to $\phi\rightarrow \infty$ corresponds instead to a BPS domain wall between the AdS minimum and the Minkowski vacuum at infinity.)

The stability of supersymmetric false vacua extends to the case 
in which the deeper minimum is not supersymmetric. In such vacuum $D_iW\neq 0$ for some $i$ and therefore $V_\mm>V_{t\mm}$: in the terminology of the previous section this is a supercritical case  and gravity quenches completely the decay. Let us illustrate
this possibility with another concrete example. Take the superpotential 
\be
W=m\Phi_1\Phi_2-\frac{\lambda}{\Lambda}(\Phi_1\Phi_2)^2\ ,
\ee
with $m$, $\lambda$ and $\Lambda$ real and positive,
and a canonical K\"ahler potential  $K=|\Phi_1|^2+|\Phi_2|^2$. 
The potential is
\bea
V&=& e^{\kappa(|\Phi_1|^2+|\Phi_2|^2)}\left\{
|\Phi_2|^2\left|m-\frac{2\lambda}{\Lambda}\Phi_1\Phi_2+
\kappa |\Phi_1|^2\left(m-\frac{\lambda}{\Lambda}\Phi_1\Phi_2\right)\right|^2
\right.\label{Vsusytononsusy}\\
&+&\left.
|\Phi_1|^2\left|m-\frac{2\lambda}{\Lambda}\Phi_1\Phi_2+
\kappa |\Phi_2|^2\left(m-\frac{\lambda}{\Lambda}\Phi_1\Phi_2\right)\right|^2-3\kappa|\Phi_1\Phi_2|^2\left|m-\frac{\lambda}{\Lambda}\Phi_1\Phi_2\right|^2
\right\} .\nonumber
\eea
Restricting the analysis to the real parts of the complex fields $\phi_i=\sqrt{2}\mathrm{Re}\Phi_i$, the potential $V(\phi_1,\phi_2)$ has a false supersymmetric vacuum at $\phi_1=\phi_2=0$ with $V_\pp=0$ and can feature two deeper supersymmetric vacua with nonzero $\phi_1$ and $\phi_2$, that transform into each other under $\phi_1\leftrightarrow \phi_2$, have $V_\mm<0$  and are degenerate.

\begin{figure}[t!]
\begin{center}
\includegraphics[width=0.35\textwidth]{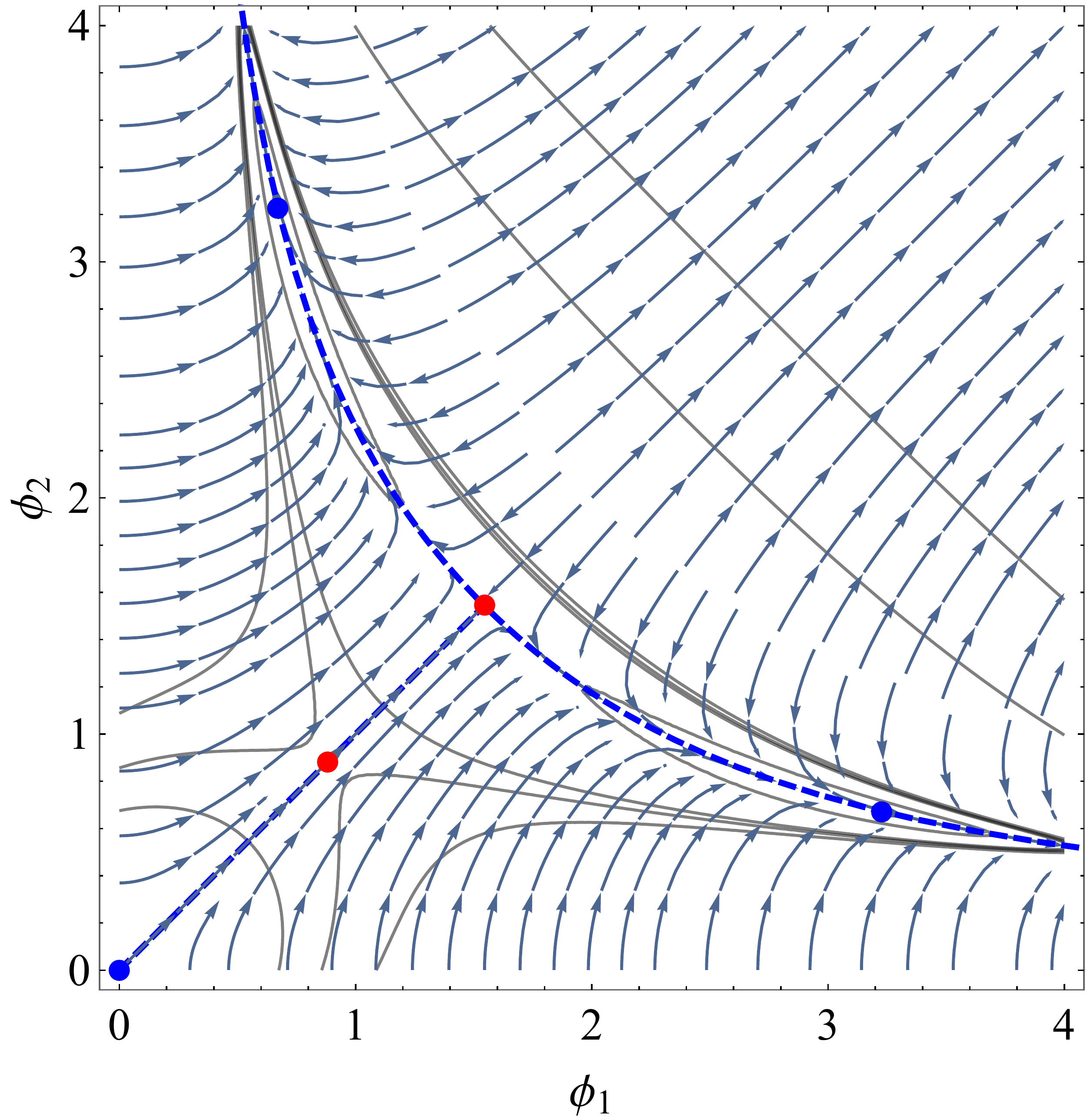}\hspace*{0.5cm}
\includegraphics[width=0.5\textwidth]{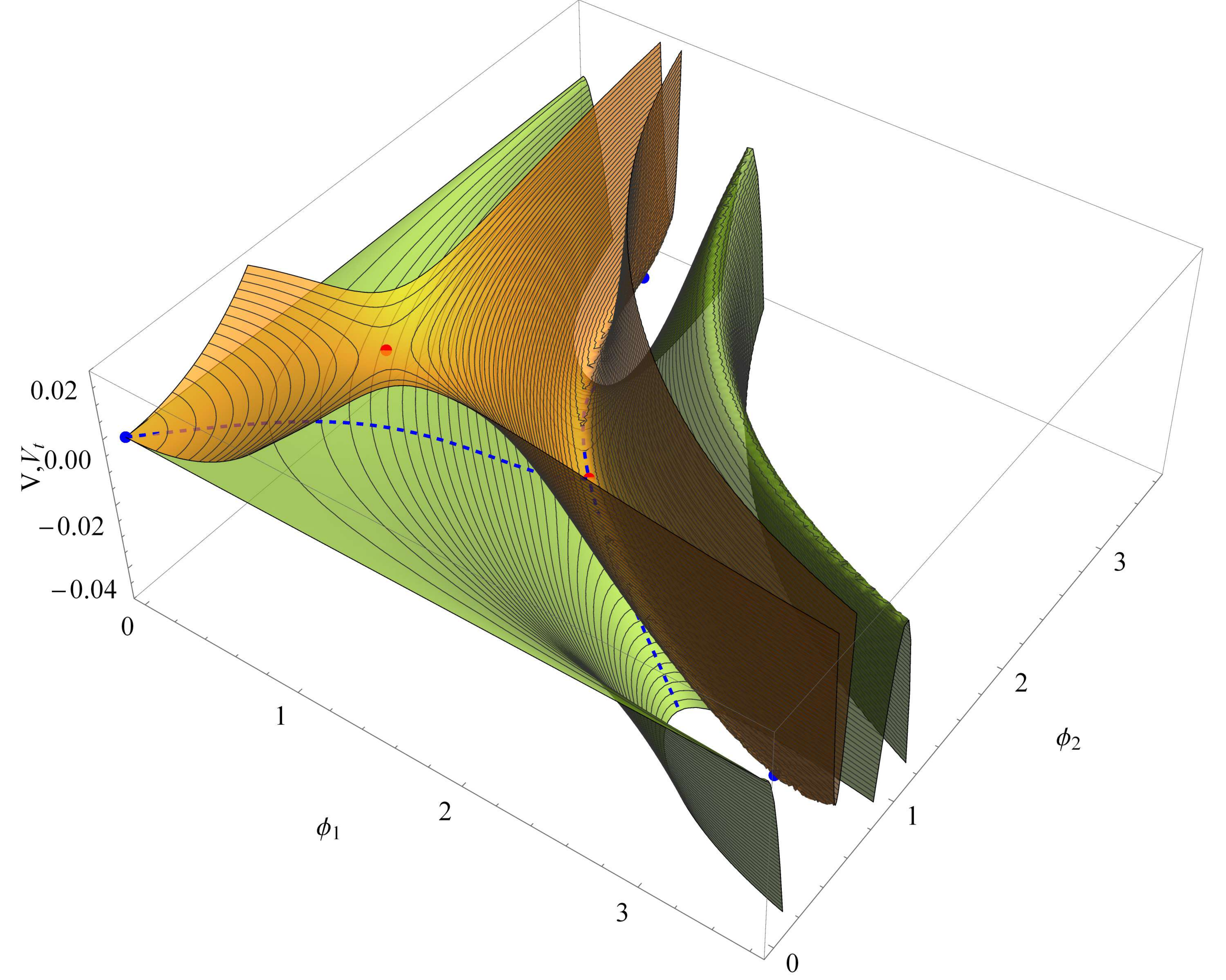}
\end{center}
\caption{For the potential (\ref{Vsusytononsusy}) with
$m=0.2$, $\kappa=0.4$, $\Lambda=10$ (in arbitrary mass units) and $\lambda=1$, the left plot gives contours of fixed $V$, with the minima marked by blue dots and the saddle points by red dots. The streamlines correspond to (\ref{SUSYpath}) and the dashed blue line to the particular streamline $\Phi(\phi)$ joining the extremals. The right plot is the 3D version of the same and shows $V$, $V_t$  and the path $\Phi(\phi)$.
\label{fig:VSUSYtoNONSUSY}
}
\end{figure}

Figure~\ref{fig:VSUSYtoNONSUSY} shows this potential for
the indicated choice of parameters. The left plot gives contour lines of $V(\phi_1,\phi_2)$, with the three minima marked by blue dots (the one at the origin and the two degenerate ones) and two saddle points in red. The streamlines shown correspond to the vector field $(d\phi_1/d\phi,d\phi_2/d\phi)$ as dictated by (\ref{SUSYpath}). A particular streamline is shown by the dashed blue line. Its straight part
connects the minimum at the origin to the (lower) saddle point, while the curved part joins that saddle point and the broken minima.
The right 3D-plot of Fig.~\ref{fig:VSUSYtoNONSUSY} shows the potential and its minima and also $V_t$, which is tangent to $V$ at the minimum and the lower saddle point but lies below the broken minima. 

To end this section, let us comment on the interpretation of the equations (\ref{SUSYpath}) as renormalization flow equations between extremals of the potential.
This interpretation, discussed in this context in \cite{CDGKL}, is inspired on the AdS/CFT duality between domain walls and renormalization group flows \cite{FGPW}. The flow equations of  
\cite{CDGKL} use as parameter to describe the flow either the radial coordinate of the domain wall or the warp factor of the metric. The tunneling potential approach followed in this paper leads in a direct way to flow equations naturally parametrized by the arc-length of the flow trajectory in field space  (\ref{SUSYpath}) . Associated to this flow one can define a 
$c$-function that changes monotonically along the flow. Originally,
this function is given by the derivative of the logarithm of the warp factor, but \cite{CDGKL} introduced an alternative $c$-function, given by $e^{\kappa K}|W|^2$. We see that this corresponds precisely (up to a proportionality constant) to the monotonic tunneling potential $V_t$ for the  supersymmetric case. 

\section{Domain Wall Profiles\label{sec:dw}}

If one is interested in studying the field profile in real space of a static domain wall between two minima in a critical case, this can be simply obtained from $V_t$ and the procedure can be applied to supersymmetric or non-supersymmetric potentials alike. Consider for simplicity a single real field case $\varphi(r)$,
with the two minima at $\varphi_\pp$ (false vacuum) and
$\varphi_\mm$ (deeper vacuum). Generically 
one has (allowing for $\varphi_\pp$ having a non-canonical kinetic term)
\be
\frac12 d\phi^2 = g(\varphi) d\varphi^2\ ,
\ee
where $\phi$ is the canonical field. Writing the metric as
$ds^2=\rho^2(r)(-dt^2+dx^2+dy^2)+dr^2$, the equations satisfied by the field $\phi(r)$ and the metric function $\rho(r)$
are simply the large $\rho$ limit of those for the Coleman-De Luccia  bounce
\bea
&&\ddot \phi+\frac{3\dot\rho}{\rho}\dot\phi = V'\ ,\label{E1}\\
&& \dot\rho^2 = \frac{\kappa}{3}\rho^2\left(\frac12\dot\phi^2-V\right)\ ,
\eea
where dots stand for derivatives with respect to $r$ and primes with respect to $\phi$.
Instead of fixing $r=0$ at the deepest minimum (like it is done for the CdL bounce) we can fix it at will, e.g. near the wall. We also choose $r$ to grow out of the false vacuum so that $\dot\phi>0$. From the basic relation between tunneling potential and Euclidean methods, $V_t=V-\dot\phi^2/2$ \cite{E}, we get then
\be
\dot\phi = \sqrt{2(V-V_t)}\ ,\quad 
\frac{\dot\rho}{\rho} =-\sqrt{\frac{-\kappa V_t}{3}}\ .
\label{dxdr}
\ee
It can be checked that, substituting $\dot\phi$ and $\dot\rho/\rho$ above on Eq.~(\ref{E1}) and using the chain rule to get $\ddot\phi=V'-V_t'$, one gets precisely the relation $D=0$.

Once $V_t$ is known, we can integrate (\ref{dxdr}), using $d\phi/d\varphi=\sqrt{2g(\varphi)}$, to get
\bea
r(\varphi) &=&\int_{-\infty}^r dr=  \int_{\varphi_\pp}^\varphi \sqrt{\frac{g(\bar\varphi)}{V(\bar\varphi)-V_t(\bar\varphi)}}\ d\bar\varphi\ ,\label{rphi}\\
\rho(\varphi) &=&  \exp\left[-\int_{\varphi_\pp}^\varphi \sqrt{\frac{-\kappa g(\bar\varphi)V_t(\bar\varphi)}{3[V(\bar\varphi)-V_t(\bar\varphi)]}}\ d\bar\varphi\right]\ ,\label{rhophi}
\eea
where we have fixed $r(\phi_\pp)=-\infty$ and $\rho(\varphi_\pp)=1$.
These expressions can be inverted numerically to get $\varphi(r)$ and then $\rho(r)$.

The profile for the Ricci curvature scalar 
\be
R=-6\left(\frac{\dot\rho^2}{\rho^2}+\frac{\ddot\rho}{\rho}\right)\ ,
\label{Rphi}
\ee
can be obtained as well, simply using (\ref{dxdr}) and its derivative
to get
\be
R(\varphi) = 6\kappa\left[V(\varphi)-\frac13 V_t(\varphi)\right]\ .
\ee

\begin{figure}[t!]
\begin{center}
\includegraphics[width=0.47\textwidth]{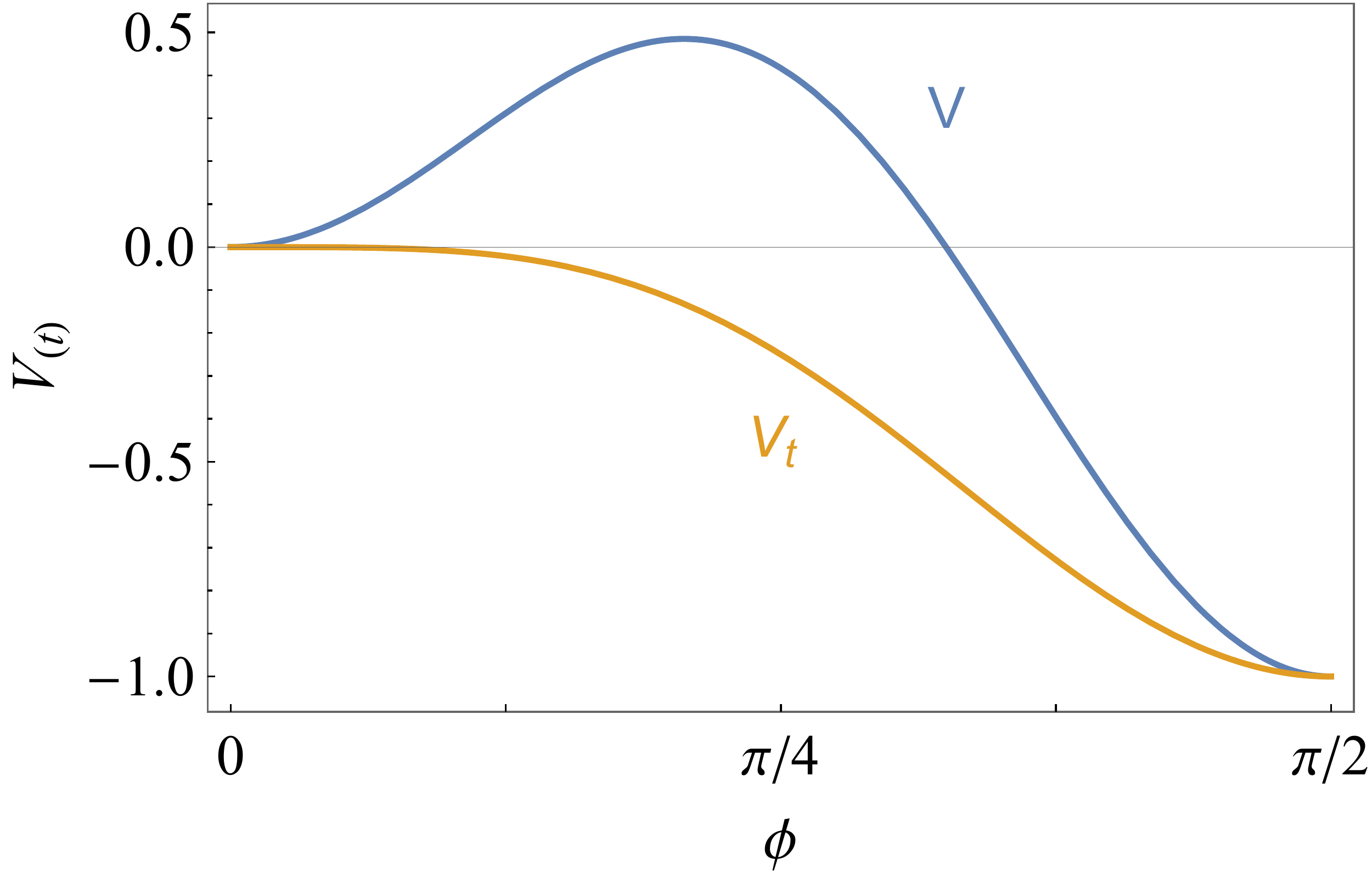}
\includegraphics[width=0.47\textwidth]{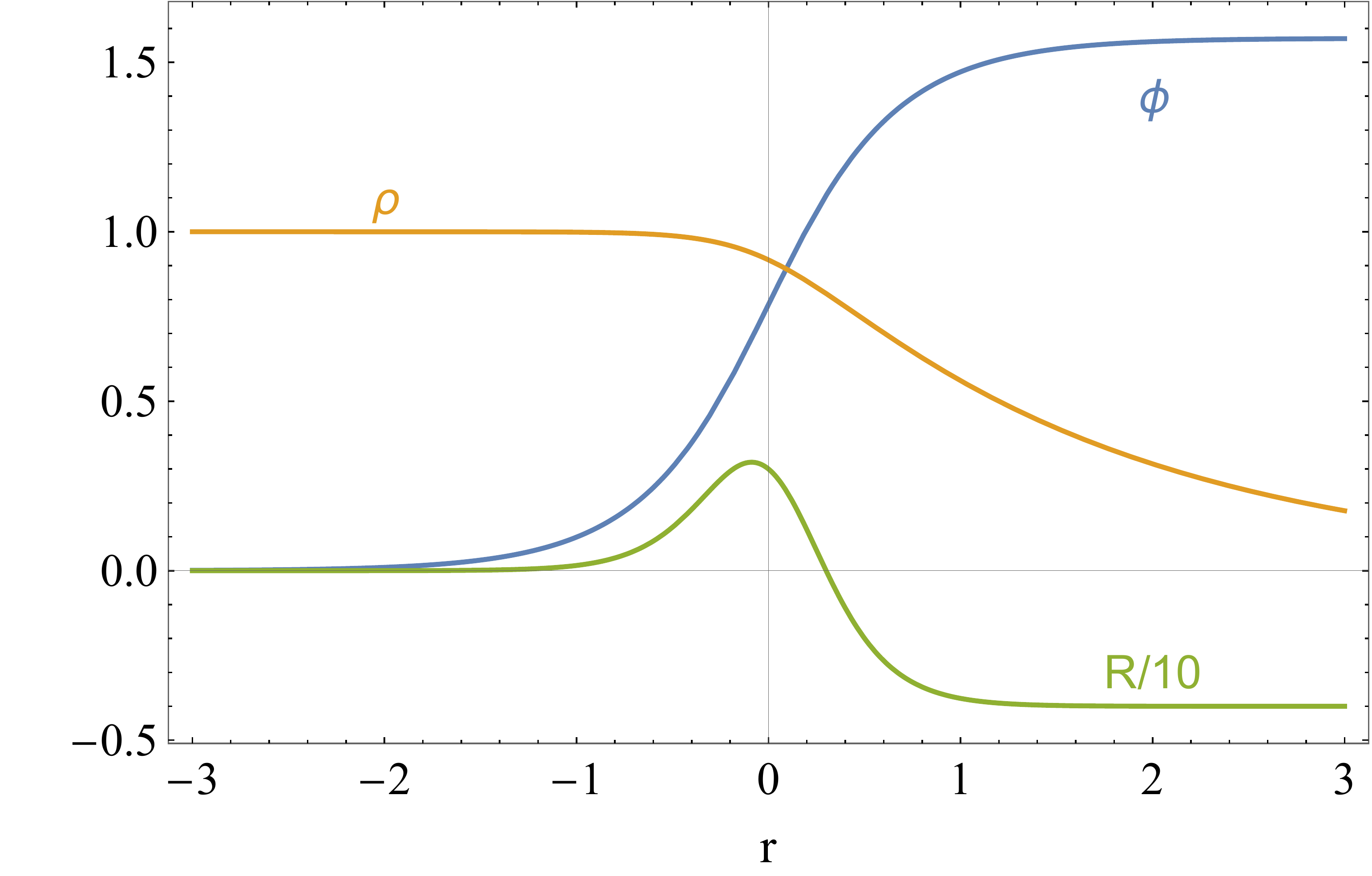}
\end{center}
\caption{Left plot: critical $V$ and $V_t$ as given in (\ref{VDW}) and (\ref{VtDW}). A static domain wall can exist between the vacua of $V$. Right plot: different profiles [field ($\phi$), metric function ($\rho$) and curvature scalar ($R$)] associated to the static domain wall. 
\label{fig:DWprofiles}
}
\end{figure}

Let us carry on the above procedure for a particular example. 
In order to be as clear as possible we use a simple analytical case, choosing
\be
V_t(\phi)=-\sin^4\phi\ .
\label{VtDW}
\ee
Using Eq.~(\ref{Vc}) we obtain the critical potential
\be
V(\phi)=\frac13\left(8-11\sin^2\phi\right)\sin^2\phi\ .
\label{VDW}
\ee
In $V$ and $V_t$ above all mass scales have been set to 1 (including $\kappa=1$).
Figure~\ref{fig:DWprofiles}, left plot, shows $V_t$ and $V$ above. The potential has a Minkowski vacuum at the origin and an AdS vacuum at $\phi=\pi/2$. With such simple potentials it is possible to obtain analytically 
the profiles for the domain wall between the two vacua of $V$.
Integrating (\ref{rphi}),  with the origin of the $r$ coordinate fixed to be at the center of the wall, with $\phi(0)=\pi/4$, and solving for $\phi$, one gets the domain wall field profile
\be
\phi(r)=\mathrm{arctan}\left(e^{4r/\sqrt{3}}\right)\ .
\label{phir}
\ee 
Integrating (\ref{rhophi}) with the boundary condition $\rho(-\infty)=1$ and using (\ref{phir}) gives the metric function profile
\be
\rho(r)=\left(1+e^{8r/\sqrt{3}}\right)^{-1/8}\ .
\ee
Finally, the Ricci curvature scalar can be obtained directly from (\ref{Rphi}) and (\ref{phir}) as
\be
R(r)=\left(1+\tanh\frac{4r}{\sqrt{3}}\right)\left(3-5\tanh\frac{4r}{\sqrt{3}}\right)\ .
\ee
 Figure~\ref{fig:DWprofiles}, right plot, shows these domain wall profiles $\phi(r), \rho(r)$ and $R(r)$ (rescaled by 1/10).  The field $\phi$ interpolates between the Minkowksi vacuum, with $\phi(-\infty)=0$, and the AdS vacuum at $\phi(\infty)=\pi/2$; the metric function $\rho(r)$ decreases away from the Minkowski vacuum; and the scalar curvature starts at zero at the Minkowski vacuum, peaks at the wall and then drops to a negative value $R=4\kappa V_\mm$ at the AdS vacuum.

\section{Near-Critical Vacuum Decay\label{sec:nearcrit}}

The tunneling potential approach is quite useful to analyze near-critical vacuum decays, allowing in particular to easily get controllable analytical examples. In this section we resort to the thin-wall approximation to study such decays, while in the next we go beyond it.

In the thin-wall limit (see \cite{Parke,Weinberg,Eg,EFT,Brown}) the wall tension of the CdL bubble is  
\be
\sigma \simeq \int_{\phi_\pp}^{\phi_\mm} \sqrt{2(V-V_t)}\ d\phi\
\simeq \
\frac{\sqrt{2}}{3\kappa}\left.\sqrt{C-6\kappa V}\right|_\pp^\mm\, ,
\ee
where $C$ is a constant, with $D^2\simeq C(V-V_t)$ \cite{Eg}. In fact, this latter approximate  equality is the thin-wall 
condition in the tunneling potential approach.
 In terms of $\sigma$,
\be
C = 6\kappa V_\pp +\frac{1}{8\sigma^2}(4\Delta V -3\kappa \sigma^2)^2\ ,
\ee
where $\Delta V\equiv V_\pp-V_\mm$. The thin-wall tunneling action
reads
\be
 S_{ tw}  = \frac{12\pi^2}{\kappa^2 V}\left.\left(1-\sqrt{1-6\kappa V/C}\right)\right|_\mm^\pp\, .
\label{tw}
\ee
In the critical case, with $D\rightarrow 0$, one has $C\rightarrow 0$
with $\sigma$ above reproducing (\ref{wtdw}) and $S_{tw}\rightarrow \infty$. Therefore, a subcritical case close to being critical is expected to correspond to small $C$ and can be analyzed in a small $C$ expansion. In that limit, (\ref{tw}) gives
\be
S_{tw}  = \frac{12\pi^2}{\kappa^2}\sqrt{\frac{6\kappa}{C}}\left(\frac{1}{\sqrt{-V_\pp}}-\frac{1}{\sqrt{-V_\mm}}\right)\, .
\ee

Let us then consider a critical potential $V_c(\phi)$ modified by a small perturbation $\epsilon\ \delta V(\phi)$ with $\epsilon\ll 1$ 
that renders the potential $V_c+\epsilon\ \delta V$ subcritical. 
This shift will induce changes in the tunneling potential and the wall tension of the CdL bubble . In an expansion in $\epsilon$:
\be
V_t=V_{tc}+\epsilon\ \delta V_t+\epsilon^2 \delta_2 V_t+{\cal O}(\epsilon^3)\ ,\quad
\sigma=\sigma_c+\epsilon\ \delta \sigma +\epsilon^2 \delta_2 \sigma +{\cal O}(\epsilon^3)\ ,
\ee
where the subindex $c$ refers to the critical values. The expansion
of $C$ gives
\bea
C &=& 6\left[\kappa \delta V_\pp+ \frac{2}{l_\pp\sigma_c}\left(\delta \Delta V_\pm-\frac{3}{l_\mm}\delta\sigma\right)\right]\epsilon\label{Cexp}\\
&+&\frac{18}{\sigma_c^2}\left\{\left[\frac13\delta\Delta V_\pm-\left(\frac{1}{l_\pp}+\frac{1}{l_\mm}\right)\delta\sigma\right]^2-\frac{1}{l_\pp l_\mm}\left[\delta\sigma^2-16\left(\frac{1}{l_\pp}-\frac{1}{l_\mm}\right)^3\frac{\delta_2\sigma}{\kappa^3\sigma_c^2}\right]\right\}\epsilon^2+{\cal O}(\epsilon^3)\ ,\nonumber
\eea
where $l_\pm\equiv \sqrt{3/(-\kappa V_\pm)}$ are the curvature lengths at the two minima. 

For the expansion of the thin-wall action, let us consider separately the cases of Minkowski or AdS vacua. When the critical potential being perturbed has a false Minkowski vacuum, the expansion (\ref{Cexp}) simplifies significantly, as $1/l_\pp=0$. 
Let us consider some examples. Take the simple critical potential of (\ref{VDW}) and modify it by a small perturbation to make it subcritical
\be
V(\phi)= V_c(\phi)+\epsilon\ \delta V =\frac13\left(8-11\sin^2\phi\right)\sin^2\phi -\epsilon \sin^2(2\phi)\ .
\ee
With $\epsilon$ small and positive, the perturbation has the effect of lowering the height of the barrier without changing the two minima
at $0$ and $\pi/2$. It is straightforward to solve for the induced change in $V_t$ by solving its EoM at first order in $\epsilon$  to get\footnote{For $\epsilon<0$ the height of the potential barrier increases and the Minkowski vacuum is made supercritical. This is reflected in a $V_t$ that is not monotonically decreasing, with $V_t\simeq -(12\epsilon /11)\phi^2$ at small $\phi$, and therefore does not describe vacuum decay.}
\be
V_t=V_{tc}+\epsilon\
\delta V_t =-\sin^4\phi- \frac{3}{11}\epsilon\sin^2(2\phi)+{\cal O}(\epsilon^2)\ .
\ee
This leads to 
\be
\sigma=\frac{2}{\sqrt{3}}\left[1-\frac{6\epsilon}{11}+{\cal O}(\epsilon^2)\right]\ ,\quad
C=-6\kappa V_\mm\frac{\delta\sigma^2}{\sigma_c^2}=\frac{216\epsilon^2}{121}+{\cal O}(\epsilon^3)\ ,
\ee
and finally to the thin-wall action
\be
S_{tw}=\frac{121\pi^2}{6\epsilon^2}+{\cal O}(1/\epsilon)\ .
\ee
The numerical factor in this result obviously depends on the function that enters in $\delta V$, which affects $\delta\sigma$. The same behavior is expected if $\delta V$
is not zero at $\phi_-$ so that the AdS minimum is downlifted.

One case of particular interest is a critical Minkowski vacuum uplifted
by a small shift to a dS one, as discussed in \cite{CDGKL}. This is the case if one chooses, for instance,
\be
\epsilon\ \delta V =\epsilon \cos^2\phi\ ,
\label{up}
\ee
in the previous example. This perturbation of $V$ shifts $V_+$ from 0 to $\epsilon$ leaving $V_-=V(\pi/2)$ unchanged. 
One can calculate analytically the shift $\delta V_t$ induced by (\ref{up}) to first order in $\epsilon$, but this is not necessary.
This case is particularly simple as the shift in $V_+$ modifies $C$ already at first order in $\epsilon$ [the change in $C$ due to the shift in $\sigma$ is ${\cal O}(\epsilon^2)$ and therefore subleading] in a simple way
\be
C = 6\kappa \delta V_\pp + {\cal O}(\epsilon^2)=6\kappa \epsilon + {\cal O}(\epsilon^2)\ ,
\ee
where we identify $\delta V_+=\epsilon$ beyond the particular example in (\ref{up}). Then it is straightforward to get
\be
S_{tw} = \frac{12\pi^2}{\kappa^2 \epsilon} +{\cal O}(1/\sqrt{\epsilon})\ ,
\ee
in agreement with the result found in \cite{CDGKL}\footnote{Notice that this action is smaller than the Hawking-Moss  action \cite{HM} by a factor 2.}. In this case the result just depends on the shift of $V_\pp$ and is independent of the shape of $\epsilon\ \delta V$. The same result holds, therefore, if the uplift is uniform, with $\delta V=\epsilon$. All that matters is that the change in $V_+$ induces a first order change in $C$. On the other hand, for $\epsilon<0$ (Minkowski vacuum downlifted to an AdS one) one gets a supercritical vacuum instead. For supersymmetric potentials, this simple case leads to $S\sim m_P^2/m_{3/2}^2$ \cite{Dine}.

For AdS to AdS decay, perturbations of order $\epsilon$ around a critical case generically give $C\sim {\cal O}(\epsilon)$, see (\ref{Cexp}), and the thin-wall tunneling action $S_{tw}\sim {\cal O}(1/\sqrt{\epsilon})$ depends on the shape of $\delta V$ via $\delta\sigma$. We do not provide here any concrete example but the next section illustrates, using a potential with AdS vacua, how a near critical case is not necessarily well described by the thin-wall approximation.

\section{Near-Critical Decays Beyond Thin-Wall\label{sec:btw}}

Although one would expect that the usual thin-wall approximation discussed in the previous section should apply to near-critical vacuum decays, we show that this is not necessarily the case using a particular example. For this we use the analytical potential presented in section~5 of \cite{Eg}. By a judicious choice of parameters we can obtain a particular potential that involves only trigonometric functions (instead of the Appell hypergeometric function $F_1$ of the general potential in \cite{Eg}). Using the notation of \cite{Eg} we choose $\kappa M^2=2$ (and measure all dimensional quantities in terms of $M=1$).
We then have
\be
V_t(\phi) = V_\pp -\mu^4 \sin^2\phi\ ,
\label{VtAdS}
\ee
and we take $V_\pp< 0$. The critical potential corresponding to this $V_t$
is 
\be
V_c(\phi)=V_t(\phi)-\frac{\mu^8\sin^2(2\phi)}{12 V_t(\phi)}\ .
\ee
Besides this critical potential, there is a subcritical potential $V(\phi)$ for which the $V_t$ above describes CdL decay. In other words, such that $V_t$ solves the corresponding EoM (\ref{EoMVt}). This subcritical potential  has the simple form
\be
V(\phi) = V_c(\phi) -\frac{\alpha(1+\alpha)\mu^{12}\cos^2\phi}{3V_t^2(\phi)\left[\left(1+\alpha\sec^2\phi\right)^{-1}+A \tan^{2\alpha}\phi\right]}\ ,
\ee
where $\alpha=V_\pp/\mu^4-1< 0$ and $A$ is a constant that can be written in terms of $\phi_0$, the exit point for the tunneling decay
out of the false vacuum at $\phi_\pp=0$. The relation is
\be
A=-\cot^{2\alpha}\phi_0\left[\frac{1}{1+\alpha\sec^2\phi_0}+\frac{\alpha(1+\alpha)\mu^4}{V_t(\phi_0)\sin^2\phi_0}\right]\ .
\label{VAdS}
\ee
For $\phi_0\rightarrow\pi/2$ one has $A\rightarrow\infty$ and $V\rightarrow V_c$ (critical limit). For other values of $\phi_0$ the potential departs from the critical potential but the tunneling potential solution describing now vacuum decay with finite action is still given by (\ref{VtAdS}), as mentioned above. This makes possible to examine analytically the behaviour of near-critical decays between the AdS minima of this potential. 

\begin{figure}[t!]
\begin{center}
\includegraphics[width=0.47\textwidth]{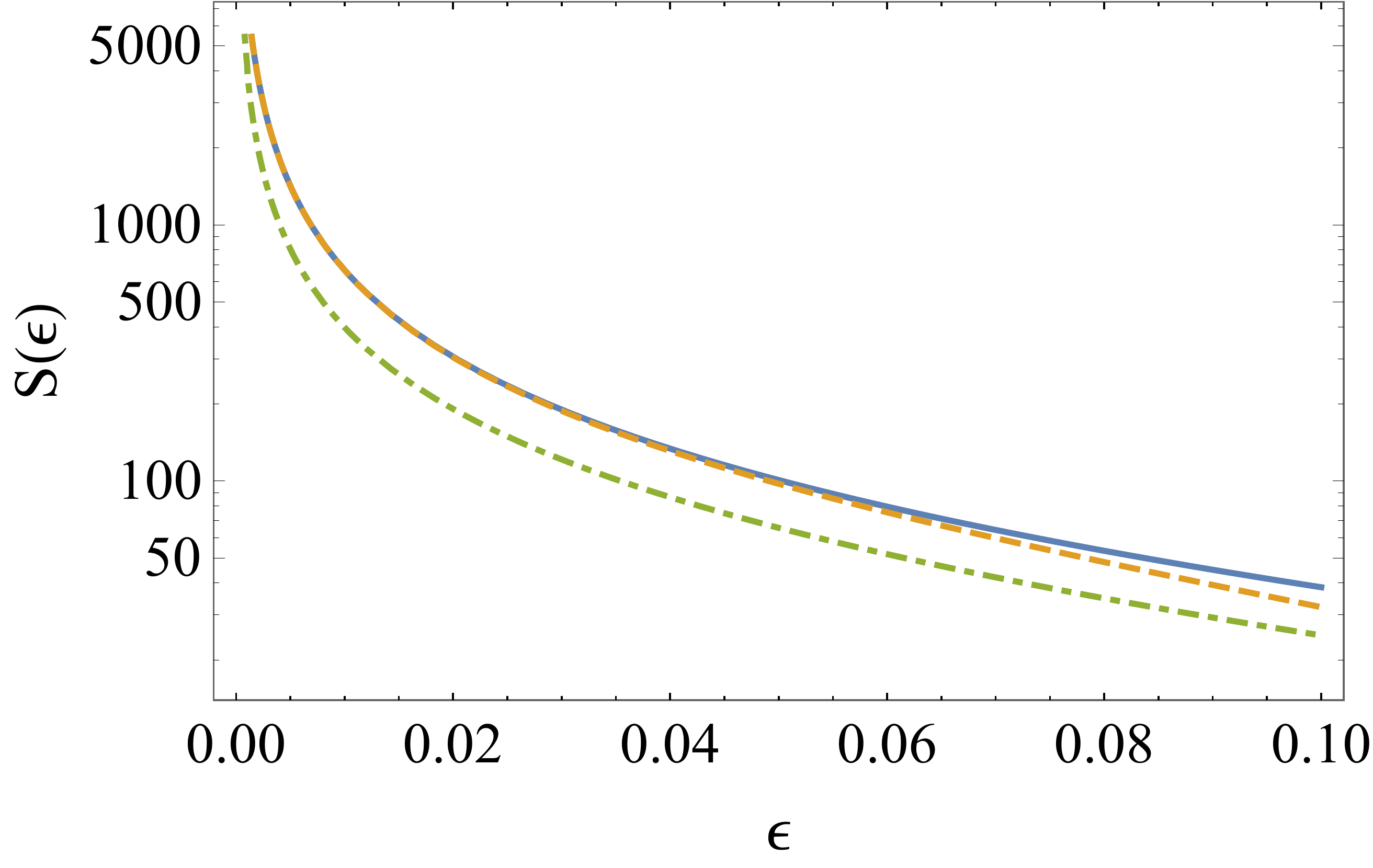}
\includegraphics[width=0.47\textwidth]{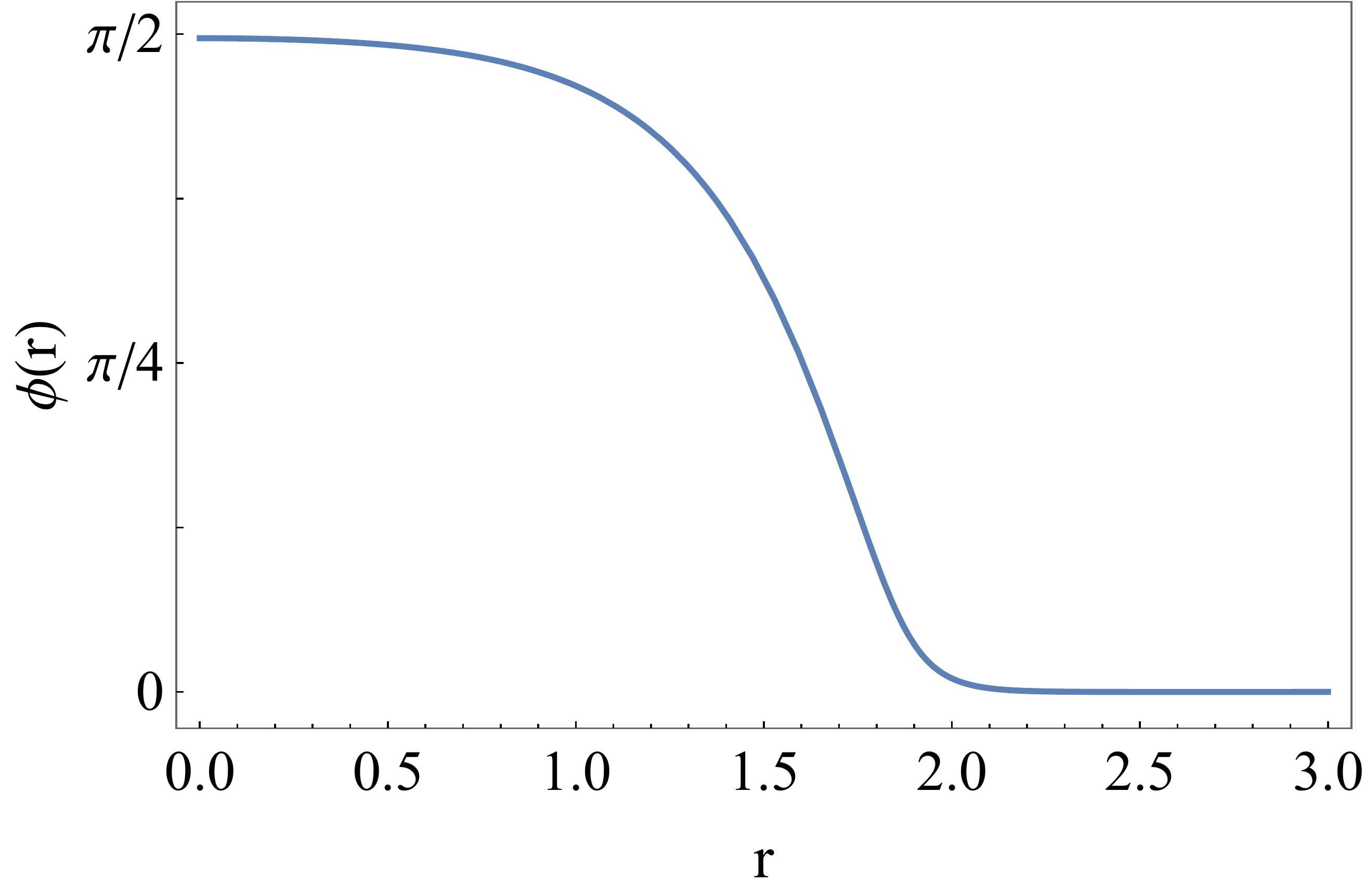}
\end{center}
\caption{Left plot: Tunneling action $S(\epsilon)$ for the potential (\ref{VAdS}) with $V_\pp=-1$, $\mu=2$, $\kappa=2$, $M=1$ and $\phi_0=\pi/2-\epsilon^{8/17}$  (solid line) compared with the analytic approximation of (\ref{Sbtw}) (dashed line) and with the thin-wall approximation with $\sigma$ computed numerically (dot-dashed line). Right plot: Euclidean bounce profile for decay of the AdS false vacuum of the previous potential, with $\epsilon=0.01$.
\label{fig:btw}
}
\end{figure}

As a concrete example take $V_\pp=-1$, $\mu=2$ and $\phi_0-\pi/2=-\epsilon^{(\mu^4/2)/(\mu^4-V_\pp)}=-\epsilon^{8/17}$. This value of $\phi_0$ is chosen so that $V-V_c\sim {\cal O}(\epsilon)$ and $D^2\sim {\cal O}(\epsilon)$. The tunneling action (\ref{SVt}) can be obtained explicitly in an expansion in small $\epsilon$ as
\bea
S(\epsilon)&=&-\frac{3\pi^{3/2}}{8\sqrt{17}\sqrt{\epsilon}}\left[\frac{\pi^{3/2}}{\cos(\pi/32)}+\frac{4}{17^{31/32}}\Gamma\left(-\frac{1}{32}\right)\Gamma\left(\frac{17}{32}\right) {}_2F_1\left(-\frac{31}{32},1;\frac{15}{32};\frac{1}{17}\right)\right]
\nonumber\\
&-&\frac{48}{17}\pi^2 +{\cal O}(\epsilon^{15/34})\ .
\label{Sbtw}
\eea
A thin-wall approximation in this case is not very precise. One gets
$\sigma =\sigma_c- {\cal O}(\epsilon^{16/17})$ and $C_{tw}={\cal O}(\epsilon^{16/17})$, leading to $S_{tw}\sim {\cal O}(1/\epsilon^{8/17})$, which is not parametrically correct. Figure~\ref{fig:btw}, left plot,
shows the tunneling action $S(\epsilon)$ (solid line) compared with the analytic approximation of (\ref{Sbtw}) (dashed line) and with the thin-wall approximation with $\sigma$ computed numerically (dot-dashed line). The bounce field profile for $\epsilon=0.01$ is shown in
the right plot of the same figure, showing that the profile indeed does not correspond to a thin-wall case. One can also check that the thin-wall condition $D^2\simeq C(V-V_t)$ is not satisfied. 

This example illustrates that, even in those near-critical cases for which the thin-wall approximation is not precise enough, alternative approximations based on the smallness of $D$ can be found.

\section{Summary and Outlook\label{sec:sumout}}

The alternative method to calculate semi-classical tunneling actions based on a variational principle formulated directly in field space \cite{E,Eg} is particularly well suited to study the impact of gravitational corrections on vacuum decay. This paper has demonstrated this in particular for the stabilizing effect of gravity for false Minkowski or AdS vacua, taking as starting point the action integral of Eq.~(\ref{SVt}). This compact expression encapsulates a plethora of different phenomena that follow quite simply and directly from it, reproducing old semi-classical results and leading to new ones, as the different sections above show.

Using that action it is straightforward to prove that gravity makes (Minkowski or AdS) false vacua more stable or that higher barriers lead to longer-lived vacua. The need to have a real $D$ in the action density of Eq.~(\ref{SVt}) plays a central  role regarding the phenomenon of gravitational quenching of vacuum decay. Examining $D$ allows one to directly formulate the rough order of magnitude conditions a vacuum/potential should satisfy to have quenching, see Eq.~(\ref{quench0}). It also allows to determine quantitatively if a  vacuum/potential is quenched or not by solving a simple first-order differential equation, see (\ref{Vtc}). Different vacua/potentials are classified as supercritical (subcritical) if vacuum decay is (not) quenched. At the frontier between both types one encounters critical cases, corresponding to $D\equiv 0$, for which vacuum decay is quenched with the CdL tunneling bubble degenerating into a BPS domain wall with tension saturating the Bogomol'nyi bound. The profile of such domain walls can also be studied starting with the tunneling potential method, see section~\ref{sec:dw}.  The general form of such critical potentials follows directly from the condition $D\equiv 0$ leading to Eq.~(\ref{Vc}). 

One case of particular interest among the critical potentials is that of supersymmetric ones, treated in Section~\ref{sec:SUSY}. It is straightforward to find the general form of the tunneling potential for supergravity potentials, Eq.~(\ref{Vtsusy}), and from it to deduce the stability of supersymmetric vacua. Based on that, one can proceed to examine the impact of supersymmetry breaking on such stability in different scenarios. More generally, one can examine near-critical vacuum decays, that is, subcritical cases that have small $D$. While the thin-wall approximation is accurate in many such cases, this is not always the case. In such instances, one can still derive approximations built as perturbative expansions with small $D$, see Section~\ref{sec:btw}.

In this paper we have not dealt with the impact of gravity on the stability of AdS maxima. This is an interesting topic that can also be analyzed with the tunneling potential approach, but it deserves a separate study all by itself and will be treated elsewhere \cite{EAdSmax}.

In closing, it is worth keeping in mind that Coleman-de Luccia vacuum decays are not the whole story, especially when dealing with vacua obtained from string theory compactifications, context in which other decay mechanisms become available.
There is in particular a recent conjecture that non-supersymmetric vacua are not stable in quantum gravity \cite{NoNonSUSY}. Originally this was based on the weak gravity conjecture \cite{WGC} applied to vacua supported by fluxes
but its validity might be more pervasive \cite{GMSV} (with non-supersymmetric vacua decaying into a bubble of nothing \cite{BoN} as generic decay mode). Even if this turns out to be true, it is still useful to be able to map whether a given non-supersymmetric vacuum is subcritical (and admits the usual Coleman-de Luccia decay with finite action) or is supercritical (and unstable only via such alternative decays). On the other hand, it might also be possible to formulate bubble of nothing decays avoiding the use of Euclidean techniques, as was done for the CdL decays with the alternative formulation used in this paper.

\appendix
\section{Multifield Noncanonical Case \label{sec:app}}
Consider a theory with scalar Lagrangian
\be
{\cal L}=\frac12 g_{i j}(\phi_k)\partial^\mu\phi^i
\partial_\mu\phi^j-V(\phi^k)\ ,
\ee
where $\phi^i$ are real fields. Assuming the potential features some false vacuum, the generic equations for the Euclidean bounce $\phi^i(r)$
describing vacuum decay (without gravity) read
\be
\ddot\phi^i+\frac{3}{r}\dot\phi^i+\Gamma_{jk}^i\dot\phi^j\dot\phi^k=g^{ik}V_{,k}\ ,
\label{multib}
\ee
where $g^{ij}$ is the inverse of the matrix $g_{ij}$, $x_{,k}\equiv dx/d\phi^k$, and
\be
\Gamma_{jk}^i\equiv \frac12 g^{il}\left(g_{kl,j}+g_{jl,k}-g_{jk,l}
\right)\ .
\ee
These equations for the bounce can be recast into a ``longitudinal'' equation (that takes the form of a single-field bounce equation) and a ``transverse'' (vectorial) equation, generalizing the canonical multi-field case 
discussed in \cite{EK}. One introduces the single field $\phi$ which is the arc length along the bounce trajectory in field space and satisfies
\be
d\phi^2 = g_{ij}d\phi^id\phi^j\ .
\ee
In terms of $\phi$, the bounce equations (\ref{multib}) split as
\bea
\ddot\phi +\frac{3}{r}\dot\phi &=& V'\ ,\label{singlebounce}\\
\dot\phi^2  D_\phi \phi^{i\prime} &= & \nabla^i_\perp V\ .
\eea
Here primes denote $\partial/\partial\phi$ and $D_\phi$ is the covariant derivative along the bounce trajectory. For the vector
$\phi^{i\prime}$ this is   
\be
D_\phi \phi^{i\prime}=\phi^{i\prime\prime}+\Gamma_{jk}^i\phi^{j\prime}\phi^{k\prime}\ .
\ee
Finally, 
\be
\nabla^i_\perp V\equiv g^{ik}V_{,k}-V'\phi^{i\prime}=V_{,k}\left(g^{ik}-\phi^{i\prime}\phi^{k\prime}\right)\ ,
\ee 
is the projection of $\vec\nabla V$ orthogonal to the bounce trajectory (so that $g_{ik}\,\phi^{i\prime}\,\nabla^k_\perp V =0$).
Adding gravity is straightforward and one arrives at the same formulas as in the single-field case for $\phi$ and the metric function $\rho(r)$, while the transverse equation is unchanged.

In the formalism of tunneling potentials, the link with the Euclidean formulation is now
\be
V_t=V-\frac12 g_{ij}\dot\phi^i\dot\phi^j =V-\frac12 \dot\phi^2\ .
\ee
The longitudinal equation reads 
\be
(4V_t'-3V')V_t'=6(V_t-V)\left[V_t''+\kappa(3V-2V_t)\right]\ ,
\ee
as in the single-field case, while the transverse equation is
\be
2(V-V_t)D_\phi \phi^{i\prime}=\nabla^i_\perp V \ .
\ee
The tunneling action $S[V_t]$ is still given by (\ref{SVt})
with the integral taken along the path parametrized by $\phi$.

For supersymmetric potentials, as those discussed in Section~\ref{sec:SUSY}, it is most natural to deal with complex fields $\Phi^i$. The equations for the Euclidean bounce in that case read
\be
\ddot\Phi^i+\frac{3}{r}\dot\Phi^i+\Gamma_{jk}^i\dot\Phi^j\dot\Phi^k=g^{i\overline k}V_{,\overline k}\ ,
\label{multibc}
\ee
where $g^{i\overline j}$ is the inverse of the K\"ahler metric $g_{\overline i j}\equiv \partial^2K/\partial\Phi^i{}^*\partial\Phi^j$, $x_{,\overline k}\equiv dx/d\Phi^{k*}$, and now
\be
\Gamma_{jk}^i\equiv g^{i\overline l} g_{k\overline l,j}\ .
\ee
Introducing the arc-length real field $\phi$ with
\be
\frac12 d\phi^2= g_{\overline i j}d\Phi^{i*}d\Phi^j\ ,
\ee
projecting (\ref{multibc}) along the bounce trajectory (multiplying by $g_{\overline m i}\Phi^{m*\prime}$) and summing with its complex conjugate, one gets back the single-field equation (\ref{singlebounce}). The transverse equation is now 
\be
2(V-V_t)D_\phi \Phi^{i\prime}=\nabla^i_\perp V \ ,
\label{TransVt}
\ee
with
\bea
D_\phi \Phi^{i\prime} &\equiv &\Phi^{i\prime\prime}+\Gamma^i_{jk}
\Phi^{j\prime}\Phi^{k\prime}\ , \nonumber\\ 
\nabla^i_\perp V & \equiv & g^{i\overline k}V_{,\overline k}-V'\Phi^{i\prime}=V_{,\overline k}\left(g^{i\overline k}-\Phi^{i\prime}\Phi^{k *\prime}\right)-V_{,k}\Phi^{k\prime}\Phi^{i\prime}\ .
\eea

For the (forbidden) tunneling between supersymmetric vacua, as discussed in Section~\ref{sec:SUSY}, one has
\bea
V&=&e^{\kappa K}\left[|DW|^2-3\kappa|W|^2\right]\ ,\nonumber\\
V_t&=&-3\kappa|W|^2e^{\kappa K}\ ,
\eea 
and the tunneling trajectory (or, more precisely, the domain wall profile)
satisfies 
\be
\Phi^{i\prime}\equiv
\frac{d\Phi^i}{d\phi} = \frac{g^{i\overline j} W \overline{D_j W}}{\sqrt{2}|W||DW|}\ .
\ee
These $V_t$ and $\Phi^{i\prime}$ satisfy the longitudinal and transverse equations discussed above. The longitudinal one, for transitions between SUSY vacua, reduces to $D=0$ (see Section~\ref{sec:SUSY}) and this is easy to check. The transverse equation can also be checked using the following results
\be
\nabla^i_\perp V = e^{\kappa K}\left\{g^{i\overline j}\left[\kappa K_{,\overline j} |DW|^2+\left(|DW|^2\right)_{,\overline j}\right]-\Phi^{i\prime}\left[\kappa K' |DW|^2+\left(|DW|^2\right)' \right]
\right\}\ ,
\ee
and
\bea
\frac{d\Phi^{i\prime}}{d\Phi^k}&=&-\Gamma^i_{lk}\Phi^{l\prime}+\kappa\delta^i_k\frac{|W|}{\sqrt{2}|DW|}+\Phi^{i\prime}\frac{W_{,k}}{2W}-\Phi^{i\prime}\frac{\left(|DW|^2\right)_{,k}}{2|DW|^2}\ ,\\
\frac{d\Phi^{i\prime}}{d\Phi^{k*}}&=&\frac{W }{\sqrt{2}|W||DW|}\left(g^{i\overline j}\overline{D_jW}\right)_{,\overline k}-\Phi^{i\prime}\frac{W_k^*}{2W^*}-\Phi^{i\prime}\frac{\left(|DW|^2\right)_{,\overline k}}{2|DW|^2}\ ,
\eea
from which (using $W'/W=W^{*\prime}/W^*$)
\be
\Phi^{i\prime\prime}=-\Gamma^i_{jk}\Phi^{j\prime}\Phi^{k\prime}
+\Phi^{i\prime}\left[\frac{\kappa |W|}{\sqrt{2}|DW|}-\frac{(|DW|^2)'}{2|DW|^2}\right]+\frac{1}{2|DW|^2}\left(g^{i\overline j}\overline{D_jW}\right)_{,\overline k}g^{\overline k l}D_lW
\ee
It is then straightforward, if tedious, to verify that (\ref{TransVt}) holds.

\section*{Acknowledgments\label{sec:ack}} 
I thank Pepe Barb\'on and Kepa Sousa for interesting discussions, and Borut Bajc, Jos\'e Juan Blanco-Pillado and Francesco Sannino for asking about the tunneling potential approach for supersymmetric vacua.
This work has been supported by the Spanish Ministry MINECO under grants  2016-78022-P and
FPA2014-55613-P and the grant SEV-2016-0597 of the Severo Ochoa excellence program of MINECO.


\begin{thebibliography}{99}

\bibitem{CdL}
  S.~Coleman, F.~De Luccia,
  Phys. Rev. D {\bf 21} (1980) 3305.

\bibitem{E}
  J.~Espinosa,
~JCAP\,{\bf 07}~(2018)~36,~\arXiv{1805.03680}{th}.
  
\bibitem{Eg}
  J.~Espinosa,
  Phys.\ Rev.\ D {\bf 100} (2019)  104007
  \arXiv{1808.00420}{th}.
    



\bibitem{ESM}
  J.~Espinosa,
  \arXiv{2003.06219}{ph}.
  
\bibitem{EK}
  J.~Espinosa and T.~Konstandin,
  JCAP {\bf 1901} (2019) 051
\arXiv{1811.09185}{th}.
    
\bibitem{GDivide}
  A.~Aguirre, T.~Banks and M.~Johnson,
  JHEP {\bf 0608} (2006) 065
  \arXivold{th/0603107}.

\bibitem{Quench2}
  D.~Samuel and W.~Hiscock,
  Phys.\ Rev.\ D {\bf 44} (1991) 3052.

  
  
\bibitem{Derrick}
  G.~Derrick,
  J.\ Math.\ Phys.\  {\bf 5} (1964) 1252.

\bibitem{EFT}
  J.~Espinosa, J.~Fortin and M.~Tr\'epanier,
  Phys.\ Rev.\ D {\bf 93} (2016) 124067
  \arXiv{1508.05343}{th}.
  
\bibitem{MPW}
  A.~Masoumi, S.~Paban and E.~Weinberg,
  Phys.\ Rev.\ D {\bf 94} (2016)  025023
\arXiv{1603.07679}{th};
Phys. Rev. D \textbf{97} (2018) 045017,
\arXiv{1711.06776}{th}.  
  
 
  
\bibitem{BFL}
  R.~Bousso, B.~Freivogel and M.~Lippert,
  Phys.\ Rev.\ D {\bf 74} (2006) 046008
  \arXivold{th/0603105}.
    
\bibitem{Boucher}
W.~Boucher,
Nucl. Phys. B \textbf{242} (1984) 282.

\bibitem{AP}
L.~Abbott and Q.~H.~Park,
Phys. Lett. B \textbf{156} (1985) 373;
Q.~H.~Park and L.~Abbott,
Phys. Lett. B \textbf{171} (1986) 223.
  
\bibitem{SWeinberg}
  S.~Weinberg,
  Phys.\ Rev.\ Lett.\  {\bf 48} (1982) 1776.
  
\bibitem{Cvetic}
  M.~Cvetic, S.~Griffies and S.-J.~Rey,
  Nucl.\ Phys.\ B {\bf 381} (1992) 301
  \arXivold{th/9201007};
  Nucl.\ Phys.\ B {\bf 389} (1993) 3
  \arXivold{th/9206004}.
 
\bibitem{SUSYvacua}
G.~Gibbons, C.~Hull and N.~Warner,
Nucl. Phys. B \textbf{218} (1983) 173,
C.~Hull,
Nucl. Phys. B \textbf{239} (1984) 541.
              
\bibitem{Positivity}
E.~Witten,
Commun. Math. Phys. \textbf{80} (1981) 381;
S.~Deser and C.~Teitelboim,
Phys. Rev. Lett. \textbf{39} (1977) 249;
C.~Hull,
Commun. Math. Phys. \textbf{90} (1983) 545.

\bibitem{BPS}
E.~Bogomolny,
Sov. J. Nucl. Phys. \textbf{24} (1976) 449;
M.~Prasad and C.~Sommerfield,
Phys. Rev. Lett. \textbf{35} (1975) 760.

\bibitem{CDGKL}
  A.~Ceresole, G.~Dall'Agata, A.~Giryavets, R.~Kallosh and A.~Linde,
  Phys.\ Rev.\ D {\bf 74} (2006) 086010
  \arXivold{th/0605266}.

\bibitem{FGPW}
D.~Freedman, S.~Gubser, K.~Pilch and N.~Warner,
Adv. Theor. Math. Phys. \textbf{3} (1999) 363,
\arXivold{th/9904017}.


\bibitem{Parke}
S.~Parke,
Phys. Lett. B \textbf{121} (1983) 313.

\bibitem{Weinberg}
  E.~Weinberg,
  ``Classical solutions in quantum field theory : Solitons and Instantons in High Energy Physics,'' Cambridge University Press, 2012.
        
\bibitem{Brown}
  A.~Brown,
  Phys.\ Rev.\ D {\bf 97} (2018) 10,  105002
  \arXiv{1711.07712}{th}.

\bibitem{HM}
  S.~Hawking and I.~Moss,
  Phys.\ Lett.\  {\bf 110B} (1982) 35.
  
\bibitem{Dine}
  M.~Dine, G.~Festuccia and A.~Morisse,
  JHEP {\bf 0909} (2009) 013
  \arXiv{0901.1169}{th}.
                        
\bibitem{EAdSmax}
J.~Espinosa, to appear.

\bibitem{NoNonSUSY}
H.~Ooguri and C.~Vafa,
Adv. Theor. Math. Phys. \textbf{21} (2017) 1787
\arXiv{1610.01533}{th};
B.~Freivogel and M.~Kleban,
\arXiv{1610.04564}{th}.

\bibitem{WGC}
N.~Arkani-Hamed, L.~Motl, A.~Nicolis and C.~Vafa,
JHEP \textbf{06} (2007) 060
\arXivold{th/0601001}.

\bibitem{GMSV}
I.~Garc\'{\i}a Etxebarria, M.~Montero, K.~Sousa and I.~Valenzuela,
\arXiv{2005.06494}{th}.

\bibitem{BoN}
E.~Witten,
Nucl. Phys. B \textbf{195} (1982) 481.
 
\end{thebibliography}
\end{document}